# Catalog of Nearby Isolated Galaxies in the Volume $z < 0.01$

I. D. Karachentsev,[1] D. I. Makarov,[1] V. E. Karachentseva,[2] and O. V. Melnyk[3,4]

[1]*Special Astrophysical Observatory of the Russian AS, Nizhnij Arkhyz 369167, Russia*

[2]*Main Astronomical Observatory, National Academy of Sciences,*

*ul. Acad. Zabolotny 27, Kiev, 03680 Ukraine*

[3]*Astronomical Observatory, Taras Shevchenko University, Observatornaya 3, Kiev, 04053 Ukraine*

[4]*Institut d'Astrophysique et de Géophysique, Université de Liège,*

*Alleé du 6 Août, 17, B5C B4000 Liege, Belgium*

We present a catalog of 520 most isolated nearby galaxies with radial velocities $V_{LG} < 3500\,\mathrm{km/s}$, covering the entire sky. This population of "space orphans" makes up 4.8% among 10 900 galaxies with measured radial velocities. We describe the isolation criterion used to select our sample, called the "Local Orphan Galaxies"(LOG), and discuss their basic optical and HI properties. A half of the LOG catalog is occupied by the Sdm, Im and Ir morphological type galaxies without a bulge. The median ratio $M_{\mathrm{gas}}/M_{\mathrm{star}}$ in the LOG galaxies exceeds 1. The distribution of the catalog galaxies on the sky looks uniform with some signatures of a weak clustering on the scale of about 0.5 Mpc. The LOG galaxies are located in the regions where the mean local density of matter is approximately 50 times lower than the mean global density. We indicate a number of LOG galaxies with distorted structures, which may be the consequence of interaction of isolated galaxies with massive dark objects.

## 1. INTRODUCTION

The population of isolated galaxies is of great interest for testing different scenarios of the origin and evolution of galaxies. Residing in the regions of very low matter density, isolated galaxies were not subjected to a significant influence from their close neighbourhood. It is assumed that over the past several billion years the evolution of these objects was driven by purely internal reasons within the "closed box" scenario. In this sense, dynamically isolated galaxies are the reference sample to study the effects of the environment on such galaxy properties as morphology, chemical abundance and the star formation rate (SFR). A recent international conference "Galaxies in Isolation: Exploring Nature Versus Nurture" held in Granada, Spain in May 2009 illustrates of a significant interest to isolated galaxies.

According to the current data [1–3], slightly more than a half of galaxies (54%) are concentrated in the virialized groups and clusters. Another 20% of galaxies are located in the collapsing regions around the groups and clusters. Over time, the population of these regions undergoes virialization as well. The remaining quarter of galaxies are referred to as the "general field galaxies", which are distributed mainly along the diffuse filaments imbordering the cosmic voids. The standard $\Lambda$CDM model of the accelerated expansion of the universe predicts that the field population will have an increasingly weakening mutual gravitational influence, and shall hence never gather in any virialized systems.

Among the fairly common category of field galaxies one can select a sample of the most isolated galaxies based on the mutual separations to their closest neighbors. A simple and effective criterion of isolation was proposed by Karachentseva [4]. A galaxy with an angular diameter of $a_1$ was considered isolated if all its significant neighbors with the angular diameters of $a_i$ in the range of

$$4 \geq a_i/a_1 \geq 1/4 \tag{1}$$

were located at the angular distances

$$X_{1i} \geq 20a_i. \tag{2}$$

At the time, in the absence of systematic data on the radial velocities and distances of galaxies, Karachentseva's criterion allowed to select in the northern sky a sample of 1 050 isolated galaxies among about 27 000 galaxies with apparent magnitudes of $m_B < 15.7^m$, which amounted to about 4% of the total. Subsequent measurements of radial velocities of the KIG galaxies [4], as well as their neighbors confirmed a good spatial isolation of these objects. Later, Karachentseva et al. [5] applied a similar criterion to search for the isolated galaxies among the extended sources of the 2MASS infrared sky survey [6]. A new 2MIG catalog covers all the northern and southern sky and contains 3 227 isolated galaxies with apparent magnitudes $K_s < 12.0^m$, and angular diameters $a_K > 30''$. The characteristic depth of the 2MIG catalog is around 6 500 km/s, and this sample contains about 6% of the galaxies with the corresponding apparent magnitudes and diameters.

For the nearby volume of space Karachentsev et al. [7] have compiled a catalog of 450 galaxies located in the sphere of a 10 Mpc radius (CNG). Most galaxies in the CNG catalog have individual distance estimates. In this most studied volume each galaxy "$i$" can be characterized by the tidal index

$$(TI)_i = \max\{\log(M_k/D_{ik}^3\} + C, \ \ k = 1, 2 \dots n, \tag{3}$$

where $M_k$ is the total mass of the neighboring galaxies, separated from the considered galaxy at the spatial distances of $D_{ik}$. The value of the constant $C$ was chosen so that the condition $TI = 0$ would correspond to the case where the Keplerian period of galaxy motion relative to its main perturbing neighbor is equal to the age of the universe $T = 13.7$ Gyr. With this definition in mind, the galaxies with the positive tidal index appear to be group members, and the condition $TI < 0$ separates the population of "field galaxies". A total of 197 galaxies, or 44% made it into the latter category. If we select the most isolated objects with $TI < -2.0$, then their relative number, i.e. about 5%, turns out to be about the same as in the KIG and 2MIG catalogs. This category of very isolated nearby galaxies settles in the regions where the local matter density (conditioned by their neighbors) is two orders lower than the mean density in the CNG catalog.

Therefore, currently there are only two samples, covering the entire sky and separating about 5% of the most isolated galaxies: a small CNG sample (with $TI < -2.0$) on the scale of 10 Mpc, and a large 2MIG sample on the scale of around 90 Mpc. The latter catalog, due to the constraint on the galaxy apparent magnitude is missing a lot of dwarf systems located at large distances. There is hence a need for a new representative sample of isolated galaxies, which is limited rather by the distance of galaxies than by the apparent magnitude. We have compiled a sample of an intermediate volume, filling the gap between the CNG and 2MIG catalogs, and present it in this paper.

## 2.  INITIAL DATA AND THE ISOLATION CRITERION

To   search for isolated galaxies we used the data on radial velocities, apparent magnitudes and morphological types from the updated HyperLEDA [8] (`http://leda.univ-lyon1.fr`) and NED (`http://nedwww.ipac.caltech.edu`) databases, complemented by the measurements of radial velocities from the latest optical and HI sky surveys: SDSS, 6dF, HIPASS, and ALFALFA. The galaxies that were checked for isolation had their radial velocities relative to the centroid of the Local Group $V_{LG} < 3\,500$ km/s at the galactic latitudes $\mid b \mid > 15°$ regardless of their luminosity and morphology. Outside this volume we as well took into account galaxies with $V_{LG} < 4\,000$ km/s and $\mid b \mid > 10°$ in order to avoid the effect of false isolation at the boundary of the sample. Each galaxy was visually examined in the DSS and SDSS images to exclude the cases of confusion in the coordinates, error identifications, and questionable radial velocities, which often arise in the implementation of automated sky surveys. In addition, we performed a morphological classification of galaxies and estimated their apparent magnitudes if they were absent in the NED and HyperLEDA databases. We used apparent magnitudes of galaxies in the $K_s$ photometric band from the 2MASS survey [6] as a basis. Apparent magnitudes in other bands ($B, V, R, I, J, H$) were transformed into the $K$-band taking into account the average color indices for each morphological type [9, 10].

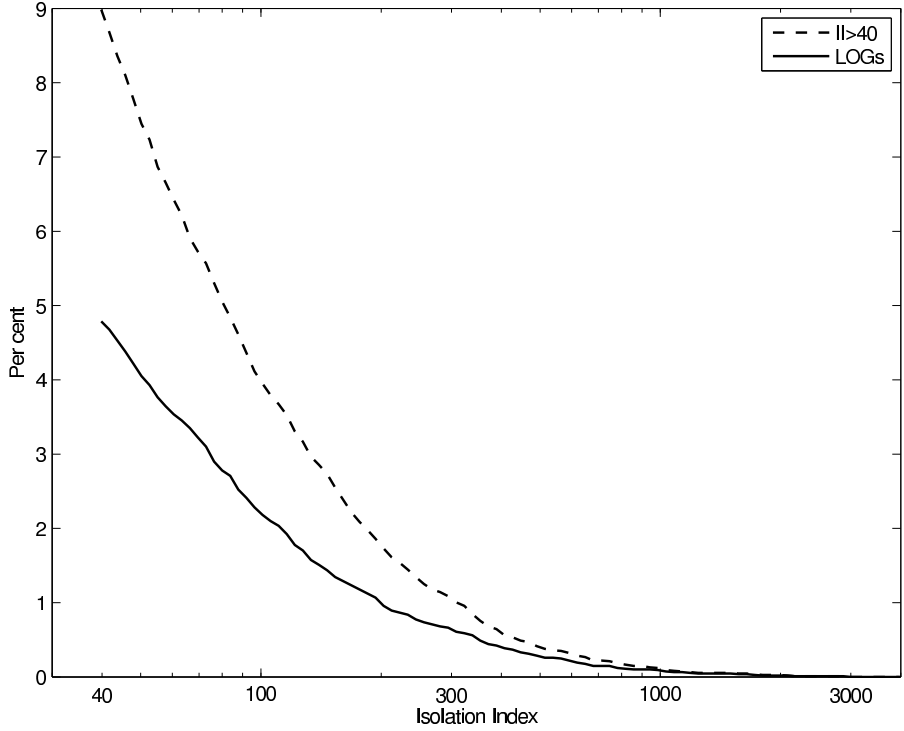

**Figure 1.** The relative number of galaxies as a function of the isolation index $II$ among 10 900 galaxies with $V_{LG} < 3500$ km/s (the dashed line). The solid line describes a similar dependence for the LOG catalog.

A total of 10 900 galaxies were checked in the above volume for galaxy system membership. Each galaxy was attributed the total mass $M$, proportional to its $K$-luminosity

$$M/L_K = \kappa M_\odot / L_\odot,$$ (4)

where $\kappa$ is a dimensionless value. A remarkable property of the $K$-luminosity is its low sensitivity to the internal absorption in a galaxy, as well as to the presence of a young blue stellar population. At $\kappa = 1$ the $K$-band luminosity corresponds to the stellar mass of a galaxy. To estimate the total mass of a galaxy we have adopted the value $\kappa = 6$, at which the structure and the known virial mass of the nearby groups is best reproduced.

Pairwise combining the galaxies into systems, we assumed that each virtual pair "$ik$" has to satisfy the condition of negative total energy

$$V_{ik}^2 R_{ik} / 2GM_{ik} < 1,$$ (5)

and the condition of its components location within the "sphere of zero-velocity" that separates the given pair against the global Hubble expansion

$$\pi H_0^2 R_{ik}^3 / 8 G M_{ik} < 1. \tag{6}$$

Here $V_{ik}$ and $R_{ik}$ denote the radial velocity differences and the projected separations of the virtual pair components, $M_{ik}$ is their total mass, expressed in terms of the $K$-luminosity at $\kappa = 6$, while $G$ and $H_0$ are the gravitational constant and the Hubble parameter, respectively. The clustering algorithm involves a sequential review of all the galaxies of the original sample and the subsequent union of bound pairs with common members into groups/clusters of galaxies.

Having applied this algorithm that takes into account individual characteristics of galaxies, about 54% of all the galaxies were combined into systems of different populations. Among the remaining population of field galaxies we then selected especially isolated ones, satisfying a higher value of the $\kappa$ parameter. To this end, we used the condition

$$\kappa = 6 \times (II), \tag{7}$$

where the dimensionless value $(II)$ has a sense of the "isolation index." At $(II) = 40$ the total of about 10% of galaxies retain their isolation. This sample of 990 galaxies formed the basis of our catalog.

The isolation criterion (5–7) is actually based on the condition of dynamical isolation of a galaxy in the 3D-space, rather than in the sky. Therefore it has a clearer physical meaning than the 2D Karachentseva's criterion (1–2). However, not all the galaxies in the volume have their radial velocities measured. To exclude the cases where the galaxy, isolated according to (5–7) may reveal a nearby neighbor in the sky with a velocity close to the velocity of the considered galaxy itself, we additionally used Karachentseva's constraint (1–2) to the already selected 990 galaxies. A consecutive use of two criteria (5–7) and (1–2) reduced our sample from 990 to 520 galaxies.

Clearly, not all the galaxies, excluded by the additional criterion (1–2) would appear to be not isolated when their "significant" neighbors in the projection would eventually have their radial velocities measured. Moreover, the results of our pilot program measuring the radial velocities of galaxies in the vicinity of the isolated galaxy candidates have shown [11] that about 80% of them retain their isolation (most of the "significant" neighbors turn out to be the distant background galaxies with a typical velocity difference of $\Delta V_{ik} > 10\,000$ km/s). Despite this, we prefer to use the more rigorously selected sample, satisfying both the (5–7) and (1–2) criteria for the further analysis.

In Fig.1, the upper (dashed) curve shows how the relative number of galaxies that satisfy the (5–6) criterion decreases with an increasing value of the isolation index $II$. For example, going from $II = 40$ to $II = 400$ the percentage of isolated galaxies drops by an order. The solid (lower) line corresponds to the case when an additional constraint is applied to the sample using the condition (1–2). At $II = 40$ the sample of isolated galaxies numbers 520, i.e., approximately the same relative number (about 5%) as in the KIG and 2MIG catalogs. We have designated the sample of these 520 galaxies as the "Local Orphan Galaxies" (LOG) and shall adhere to this acronym further down.

An example of a situation where new redshift measurements can break the isolation of the galaxy, selected by the (5–7) criterion, is LOG 227. Near the galaxy KUG 0956+420 with velocity $V_h = 1\,682$ km/s there was discovered a neighboring galaxy KUG 0956+419, the radial velocity of which $V_h = 1\,737$ km/s was measured with a great error ($\pm 340$ km/s). Both blue dwarf galaxies may be forming a physical pair with a projection distance of 40 kpc.

## 3. THE LOG CATALOG

A list of 520 "orphan"galaxies in the Local Supercluster and its surroundings is presented in Table 1. Its columns contain the following data: (1) is the running number in the catalog, (2) is the name or the number of a given galaxy in the known catalogs as they are fixed in the HyperLEDA and NED databases; in some cases the long names of galaxies from the past surveys (SDSS, 6dF, HIPASS) are listed with an ellipsis and with no coordinate part, (3)—equatorial coordinates for epoch (2000.0), (4)—radial velocity of the galaxy relative to the centroid of the Local group and its measurement error (in km/s), (5)—morphological type of the galaxy in the digital de Vaucouleurs scale [12], (6)—integrated apparent magnitude of the galaxy in the $K_s$-band, adopted from the 2MASS survey or converted from other photometric bands taking into account the morphological type; in the latter case, which dominates, the $K$ magnitude error may reach about $0.5^m$; (7)—the value of the isolation index in the condition $M/L_K = 6 \times (II) \times (M_\odot/L_\odot)$, at which the galaxy still retains its isolation, (8)—the logarithm of the flux in the HI 21 cm line (in Jansky/km/s) according to the HyperLEDA; in some cases (marked by a colon), the upper limit of the HI-flux was estimated according to the HIPASS survey data or other available sources, (9)—in this column "+" marks the galaxies satisfying Karachentseva's criterion [4] with a significant margin, (10)—the comment column contains a brief reference to the following data: "IR" indicates the presence in a galaxy of an infrared flux in the IRAS survey, "F" marks the belonging of a galaxy to the flat systems from the RFGC and 2MFGC catalogs, "pec" marks the presence of a peculiar structure, "Mrk" indicates the belonging of a galaxy to the list of Markarian active objects, and "KIG"—to the isolated galaxies from the KIG catalog.

## 4. BASIC CHARACTERISTICS OF LOG GALAXIES

Figure 2 presents a distribution of isolated galaxies in the radial velocity intervals of 250 km/s. The unshaded histogram shows the total number of galaxies satisfying the $II > 40$ criterion, and the shaded part corresponds to the LOG catalog galaxies, for which the additional isolation condition (1–2) is applied. The diamonds mark the percentage of LOG galaxies to the total number of galaxies in each velocity interval (the right-hand scale). As we can see, the relative number of isolated galaxies is practically independent of the distance, undergoing the statistical fluctuations around the mean value of 4.8%. An approximate constancy of the LOG galaxy percentage indicates that the isolation criterion we used works equally effectively both in the nearby and distant volumes. This circumstance is not trivial. For example, in a sample of isolated galaxies from the SDSS survey [13] their relative number varies with distance more than tenfold.

The overall distribution of 520 LOG galaxies in the sky in equatorial coordinates is demonstrated in Figure 3. The galaxies from the closer volume with $V_{LG} < 2000$ km/s are marked by the circles of a larger diameter. The region of a significant galactic absorption along the Milky Way is shown by the ragged gray stripe. The distribution of isolated galaxies outside the region $\mid b \mid < 15°$ looks pretty uniform. Some deficiency of isolated objects is noticeable in the direction of the known nearby clusters Virgo, Fornax and similar groups around M81 and CenA.

An important characteristic of each galaxy sample is its morphological composition, which retains the features of the evolution of galaxies. Figure 4 shows the distribution of 520 LOG galaxies, as well as all the 990 isolated galaxies with $II > 40$ according to the morphological type. As follows from this histogram, the use of an additional isolation condition (1–2) does not introduce any special selectivity according to the morphological type. The latest types of galaxies T = 8–10, i.e. Sdm, Im, Ir are the most common members of the LOG sample. They account for 51% of all galaxies.

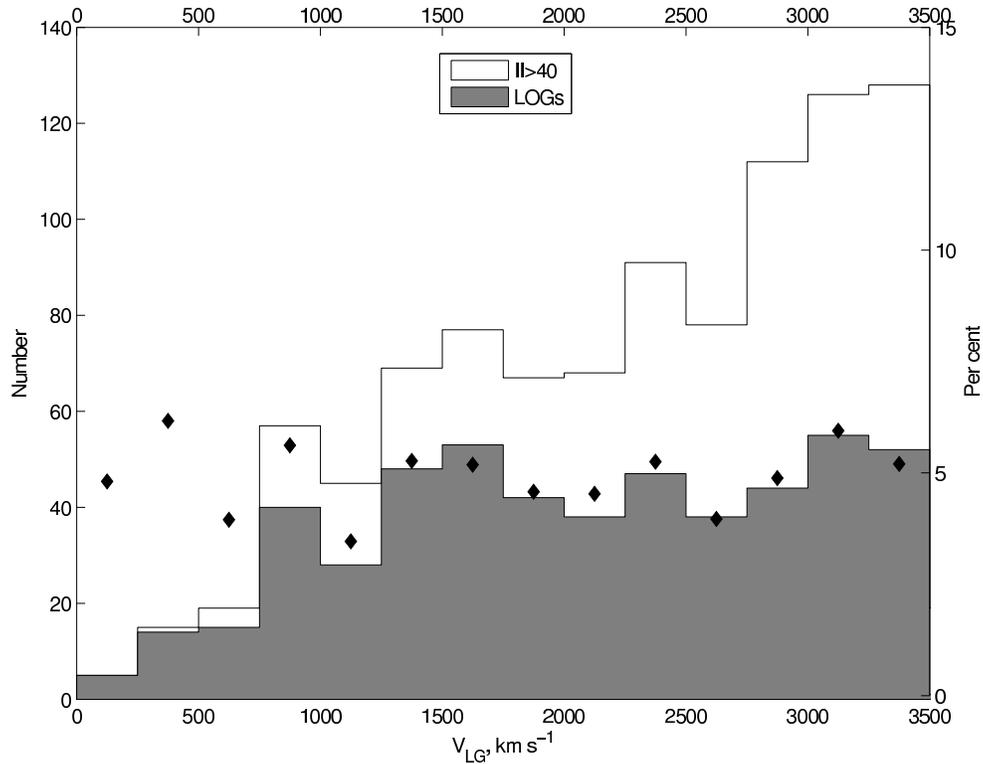

**Figure 2.** The distribution of galaxies with the $II > 40$ isolation index (the unshaded histogram) and LOG galaxies (the gray histogram) by radial velocity. The diamonds indicate the relative number of LOG galaxies in each velocity interval (percentage of the total number N = 10 900, the right-hand scale).

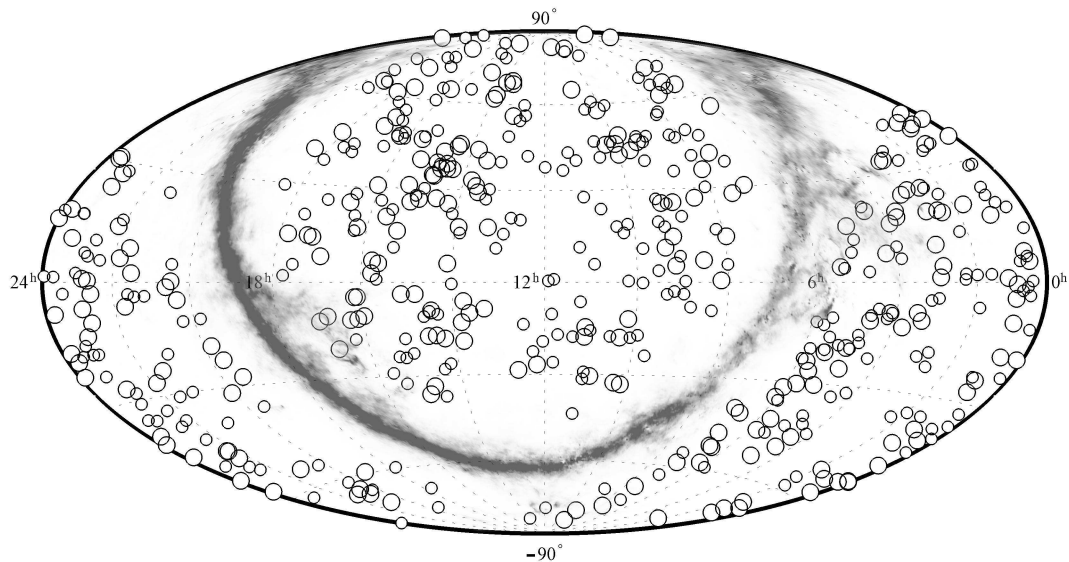

**Figure 3.** The distribution of 520 LOG galaxies on the sky in equatorial coordinates. The galaxies with radial velocities $V_{LG} \leq 2\,000$ km/s are marked by larger circles.

**Table 1.** Catalog of isolated galaxies in the Local Supercluster and its neighborhood

| LOG | Name | RA (J2000.0) Dec. | $V_{LG}$ | ± | $T$ | $K_s$ | lg $II$ | lg $F_{HI}$ | K73 | Note |
|---|---|---|---|---|---|---|---|---|---|---|
| (1) | (2) | (3) | (4) | (5) | (6) | (7) | (8) | (9) | | (10) |
| 1 | ESO149-013 | 000246.3−524618 | 1363 | 80 | 10 | 13.39 | 1.78 | 1.06 | + | |
| 2 | ESO149-018 | 000714.5−523712 | 1744 | 9 | 9 | 13.61 | 1.86 | 0.74 | + | |
| 3 | UGC00064 | 000744.0+405232 | 554 | 17 | 10 | 12.80 | 1.97 | 1.24 | + | |
| 4 | UGC00063 | 000750.8+355759 | 715 | 5 | 10 | 12.72 | 2.03 | 0.36 | + | |
| 5 | ESO538-024 | 001017.8−181551 | 1634 | 5 | 10 | 13.20 | 1.87 | 0.84 | + | |
| 6 | PGC130903 | 001108.7−385915 | 3180 | 43 | 6 | 13.95 | 2.17 | 0.3 : | | |
| 7 | 6dF... | 001408.3−353648 | 3268 | 29 | 9 | 14.31 | 2.25 | 0.71 | + | |
| 8 | SDSS... | 001500.1−110804 | 3467 | 5 | 6 | 14.78 | 1.92 | 0.3 : | + | |
| 9 | ESO241-027 | 001502.7−431731 | 3235 | 74 | 6 | 12.44 | 1.99 | 0.05 | + | |
| 10 | 6dF... | 001550.9−225511 | 3213 | 34 | 6 | 12.54 | 2.07 | 0.56 | + | |
| 11 | ESO194-002 | 001830.4−473921 | 1433 | 10 | 9 | 13.92 | 1.69 | 0.3 : | + | |
| 12 | AM0016-575 | 001909.3−573830 | 1636 | 5 | 2 | 12.12 | 1.87 | 1.28 | + | pec |
| 13 | UGC00199 | 002051.8+125122 | 2015 | 5 | 10 | 14.64 | 1.72 | 0.53 | + | |
| 14 | ESO150-005 | 002225.6−533851 | 1344 | 5 | 8 | 11.32 | 2.02 | 1.11 | + | |
| 15 | NGC0101 | 002354.6−323210 | 3411 | 30 | 6 | 10.19 | 2.01 | 1.12 | + | IR |
| 16 | UM240 | 002507.4+001846 | 3397 | 13 | 9 | 15.64 | 2.30 | 0.3 : | + | |
| 17 | 6dF... | 002755.3−031101 | 3372 | 10 | 6 | 11.79 | 2.05 | 0.46 | + | |
| 18 | UM040 | 002826.6+050016 | 1523 | 5 | 2 | 13.26 | 1.96 | 0.70 | + | |
| 19 | UGC00285 | 002851.1+285622 | 2428 | 11 | 2 | 11.60 | 2.08 | 0.04 | + | IR |
| 20 | UGC00288 | 002903.6+432554 | 463 | 5 | 10 | 13.32 | 1.85 | 0.71 | + | |
| 21 | UGC00313 | 003126.1+061224 | 2237 | 35 | 4 | 11.33 | 2.12 | −0.09 | + | IR |
| 22 | HS0029+1748 | 003203.1+180446 | 2410 | 30 | 9 | 15.12 | 2.53 | 0.3 : | | |
| 23 | ESO294-020 | 003209.7−401605 | 1393 | 11 | 8 | 11.13 | 1.90 | 0.44 | + | IR |
| 24 | UGC00328 | 003322.1−010717 | 2139 | 5 | 8 | 12.78 | 1.97 | 1.11 | + | |
| 25 | CGCG409-040 | 003448.9+072701 | 739 | 10 | 2 | 11.44 | 1.76 | −0.58 | + | |
| 26 | CGCG500-052 | 003707.6+284954 | 2233 | 16 | 9 | 12.42 | 2.08 | 0.64 | | F |
| 27 | PGC002235 | 003726.1−370311 | 3400 | 24 | 6 | 14.07 | 2.37 | 0.3 : | | |
| 28 | HIPASSJ0041-01b | 004139.6−020042 | 2096 | 9 | 10 | 13.49 | 2.01 | 0.52 | + | |
| 29 | ESO540-016 | 004214.7−180942 | 1623 | 5 | 6 | 12.50 | 1.67 | 1.52 | + | IR,F |
| 30 | Andromeda IV | 004230.1+403433 | 522 | 9 | 10 | 13.98 | 1.77 | 1.36 | + | |
| 31 | CGCG410-002 | 004448.4+050809 | 3066 | 21 | 10 | 11.66 | 1.94 | 0.3 : | + | IR |
| 32 | UGC00477 | 004613.1+192924 | 2870 | 10 | 7 | 11.81 | 2.11 | 1.59 | | F |
| 33 | UGCA014 | 004747.5−095358 | 1450 | 5 | 7 | 11.93 | 2.00 | 1.40 | | F |
| 34 | ESO079-007 | 005003.8−663312 | 1511 | 11 | 4 | 11.67 | 2.00 | 0.84 | | IR |
| 35 | PGC169954 | 005154.6+232851 | 2847 | 8 | 9 | 13.90 | 1.88 | 0.01 | | |
| 36 | ARK018 | 005159.6−002912 | 1764 | 8 | 1 | 12.58 | 1.89 | 1.17 | | |
| 37 | MCG-01-03-027 | 005217.2−035760 | 1520 | 28 | 6 | 13.93 | 1.85 | 1.05 | + | |
| 38 | ESO411-027 | 005251.7−271933 | 1852 | 10 | 8 | 13.43 | 1.95 | 0.27 | | |
| 39 | IC1596 | 005442.8+213122 | 2891 | 5 | 3 | 11.41 | 1.61 | 0.70 | + | IR,KIG |
| 40 | UGC00578 | 005621.1+394933 | 1728 | 31 | 4 | 11.46 | 2.32 | | | IR |
| 41 | ESO474-045 | 005721.3−242219 | 1894 | 5 | 8 | 13.73 | 2.03 | 0.64 | + | |
| 42 | UGC00614 | 005936.2+353337 | 2591 | 13 | 6 | 11.98 | 2.74 | 0.67 | + | IR |
| 43 | ESO151-019 | 010220.8−541923 | 1261 | 9 | 8 | 11.81 | 2.15 | 0.86 | + | |
| 44 | UGC00655 | 010401.2+415035 | 1084 | 5 | 8 | 12.57 | 1.63 | 1.21 | + | |
| 45 | MCG-04-03-052 | 010632.4−234045 | 3500 | 33 | 5 | 13.25 | 2.06 | 0.54 | + | |
| 46 | UGC00685 | 010722.4+164104 | 351 | 5 | 9 | 12.00 | 2.24 | 1.06 | + | KIG |
| 47 | NGC0406 | 010725.1−695245 | 1335 | 5 | 5 | 10.17 | 1.63 | 1.53 | + | IR |
| 48 | UGC00695 | 010746.4+010349 | 764 | 5 | 9 | 12.76 | 2.21 | 0.53 | | |
| 49 | KK11 | 010821.9−381234 | 605 | 10 | 9 | 14.26 | 2.05 | 0.67 | + | |
| 50 | NGC0404 | 010927.0+354304 | 218 | 24 | −3 | 8.55 | 1.78 | 1.59 | + | IR |
| 51 | ESO243-050 | 011048.8−422231 | 1423 | 8 | 10 | 12.12 | 2.03 | 0.84 | + | |
| 52 | AM0117-681 | 011901.8−680244 | 1883 | 9 | 7 | 12.69 | 2.13 | 0.72 | + | |
| 53 | LSBGF352-021 | 012658.6−350542 | 2068 | 9 | 10 | 15.26 | 1.98 | 0.16 | | |
| 54 | AM0126-653 | 012822.4−651615 | 1461 | 22 | 5 | 11.55 | 1.91 | 0.95 | + | IR |
| 55 | UGC01054 | 012848.3+342047 | 2889 | 5 | 7 | 14.25 | 2.60 | 0.86 | | F |
| 56 | FGC0175 | 013531.5+020154 | 2731 | 5 | 8 | 13.65 | 1.92 | 0.65 | + | F |
| 57 | NGC0620 | 013659.7+421924 | 2708 | 19 | 7 | 10.99 | 1.94 | 2.00 | + | IR |

| (1) | (2) | (3) | (4) | (5) | (6) | (7) | (8) | (9) | (10) |
|---|---|---|---|---|---|---|---|---|---|
| 58 | UGCA020 | 014314.7+195832 | 677 | 5 | 10 | 13.61 | 1.75 | 0.97 | + | pec |
| 59 | ESO114-007 | 014630.4−584027 | 2063 | 9 | 8 | 11.98 | 2.14 | 1.22 | + | |
| 60 | UGC01207 | 014726.3+820934 | 1500 | 5 | 8 | 14.01 | 1.65 | 0.75 | | |
| 61 | NGC0685 | 014742.8−524543 | 1234 | 16 | 5 | 9.18 | 1.62 | 1.54 | + | IR |
| 62 | UGC01198 | 014917.7+851538 | 1408 | 38 | −2 | 10.89 | 1.81 | 0.02 | + | IR |
| 63 | KUG0148-067 | 015128.4−063060 | 2205 | 5 | 7 | 12.40 | 1.93 | 1.03 | + | |
| 64 | ESO354-021 | 015833.3−345147 | 3046 | 43 | 4 | 13.12 | 1.79 | 0.54 | + | IR |
| 65 | UGC01464 | 015907.6+015324 | 3049 | 75 | 8 | 14.01 | 2.29 | 0.3 : | + | |
| 66 | ESO153-019 | 020539.2−560411 | 2014 | 30 | 6 | 12.47 | 1.96 | 0.32 | + | |
| 67 | ESO030-008 | 020901.7−755606 | 1054 | 60 | 7 | 12.46 | 1.88 | 1.39 | + | F |
| 68 | PGC138508 | 021223.2+131227 | 2375 | 27 | 8 | 13.80 | 2.41 | 0.00 | | |
| 69 | PGC169963 | 021336.1+231528 | 2617 | 8 | 8 | 14.78 | 2.71 | −0.05 | | |
| 70 | UGC01756 | 021653.9+021212 | 3098 | 6 | 2 | 11.67 | 1.98 | 0.3 : | + | IR |
| 71 | KUG0215+005 | 021808.1+004530 | 2820 | 5 | 8 | 14.72 | 1.88 | 0.3 : | + | |
| 72 | ESO355-005 | 021839.7−363152 | 2399 | 30 | 8 | 14.68 | 1.64 | 0.57 | | |
| 73 | NGC0918 | 022550.8+182946 | 1648 | 5 | 5 | 8.93 | 2.08 | 1.22 | + | IR,KIG |
| 74 | UGC01970 | 022954.0+251523 | 2073 | 5 | 6 | 11.14 | 1.71 | 1.02 | + | F |
| 75 | UGC01975 | 023014.7+330757 | 3369 | 14 | 5 | 11.95 | 1.88 | 0.48 | + | F,KIG |
| 76 | UGC01999 | 023152.6+190911 | 1110 | 5 | 5 | 12.92 | 2.03 | 1.32 | + | F |
| 77 | UGC02082 | 023616.1+252526 | 862 | 6 | 6 | 9.97 | 1.82 | 1.66 | | IR,F,KIG |
| 78 | ESO479-020 | 023908.2−223944 | 2999 | 11 | 5 | 12.30 | 1.84 | 0.99 | + | |
| 79 | UGC02143 | 023936.6+360452 | 2930 | 9 | 2 | 11.27 | 1.83 | 0.64 | | IR |
| 80 | ESO546-021 | 024742.8−185025 | 3037 | 5 | 7 | 13.87 | 1.66 | 0.54 | | F |
| 81 | CGCG463-025 | 025019.7+190643 | 1362 | 5 | 9 | 12.27 | 2.67 | 0.47 | | |
| 82 | UGC02352 | 025205.4+042215 | 1872 | 8 | 7 | 14.40 | 2.58 | 0.68 | + | F |
| 83 | UGC02392 | 025546.4+334560 | 1712 | 5 | 5 | 13.22 | 2.39 | 0.78 | + | |
| 84 | UGC02429 | 025709.2+011935 | 1822 | 6 | 7 | 12.88 | 2.19 | 0.84 | + | |
| 85 | UGC02432 | 025726.8+100812 | 840 | 5 | 10 | 13.84 | 2.57 | 0.85 | + | |
| 86 | UGCA047 | 025824.3−041744 | 2397 | 10 | 6 | 11.37 | 1.61 | 0.85 | + | IR |
| 87 | NGC1156 | 025942.2+251414 | 507 | 5 | 9 | 9.46 | 2.12 | 1.67 | + | IR,KIG |
| 88 | KKH018 | 030305.8+334140 | 375 | 5 | 10 | 13.49 | 1.76 | 0.40 | + | |
| 89 | ESO547-004 | 030330.6−201034 | 3275 | 10 | 5 | 13.29 | 1.99 | 0.64 | + | |
| 90 | MCG-01-09-010 | 030931.2−045444 | 3136 | 5 | 6 | 11.29 | 1.96 | 0.77 | | |
| 91 | MCG+04-08-013 | 031419.8+240914 | 1421 | 9 | 6 | 13.44 | 2.61 | 1.05 | | |
| 92 | KDG032 | 031856.7−103247 | 1910 | 9 | 10 | 14.02 | 1.61 | 0.84 | + | |
| 93 | NGC1337 | 032806.0−082319 | 1216 | 7 | 6 | 9.19 | 1.87 | 1.80 | + | IR |
| 94 | UGC02809 | 033933.2+194703 | 1368 | 6 | 8 | 12.59 | 1.60 | 0.77 | + | |
| 95 | HIPASSJ0341+18 | 034201.8+180830 | 1369 | 9 | 8 | 12.89 | 1.60 | 0.48 | | |
| 96 | 2MASX... | 034559.4−123149 | 900 | 25 | 1 | 11.17 | 2.28 | 0.3 : | + | IR |
| 97 | UGC02899 | 035424.6+063524 | 3482 | 5 | 5 | 12.14 | 2.11 | 1.01 | + | F |
| 98 | UGC02905 | 035700.1+163121 | 345 | 5 | 10 | 13.60 | 2.19 | 0.66 | | |
| 99 | HIPASSJ0358+10 | 035824.2+095845 | 2003 | 9 | 8 | 13.70 | 2.82 | 0.84 | + | |
| 100 | MCG-01-11-002 | 040615.9−083809 | 2777 | 5 | 8 | 13.72 | 1.75 | 0.74 | + | pec |
| 101 | PGC103224 | 040755.9−444750 | 3380 | 31 | 7 | 14.73 | 2.38 | 0.15 | + | |
| 102 | [KKS2000]53 | 040904.1−083737 | 834 | 5 | 9 | 12.51 | 2.08 | 0.77 | | |
| 103 | 2MASX... | 041202.3−101841 | 2815 | 74 | 4 | 12.70 | 2.60 | 0.3 : | | IR |
| 104 | ESO420-013 | 041349.7−320025 | 3449 | 36 | −1 | 9.47 | 1.79 | 0.3 : | + | IR |
| 105 | UGC02997 | 041604.9+081049 | 1590 | 19 | 2 | 10.19 | 2.05 | −0.03 | + | IR |
| 106 | 6dF... | 041804.5−214740 | 3468 | 74 | 6 | 13.52 | 2.55 | 0.3 : | + | |
| 107 | UGC03045 | 042641.7+202428 | 1429 | 5 | 5 | 9.58 | 2.48 | 0.3 : | | F |
| 108 | HIPASSJ0426-07 | 042647.9−073332 | 2368 | 9 | 8 | 14.76 | 1.92 | 0.64 | + | |
| 109 | UGC03053 | 042809.7+213919 | 2452 | 5 | 6 | 9.71 | 2.24 | 0.75 | + | IR |
| 110 | ESO251-003 | 042841.3−461916 | 1194 | 9 | 9 | 13.90 | 1.72 | 0.57 | | |
| 111 | ESO202-035 | 043216.4−494033 | 1663 | 18 | 5 | 10.03 | 1.90 | 1.58 | + | IR,F |
| 112 | 2MASX... | 043342.0−333046 | 2677 | 64 | 1 | 11.03 | 2.54 | 0.3 : | | |
| 113 | ESO304-002 | 043519.8−421212 | 2030 | 51 | 2 | 11.84 | 2.35 | 0.3 : | + | IR |
| 114 | MCG-02-12-046 | 043624.6−093049 | 2327 | 7 | 8 | 12.70 | 1.84 | 1.13 | + | IR,pec |
| 115 | PGC971141 | 043632.5−110009 | 1727 | 17 | 8 | 13.78 | 2.25 | 0.3 : | + | |
| 116 | ESO551-030 | 044226.8−172713 | 3067 | 30 | 7 | 12.39 | 1.83 | 0.75 | + | |
| 117 | KKH028 | 044344.0+025954 | 3463 | 5 | 10 | 13.49 | 2.36 | 0.55 | + | |
| 118 | APMUKS... | 044720.7−151518 | 2375 | 49 | 10 | 14.55 | 1.92 | 0.31 | | |

| (1) | (2) | (3) | (4) | | (5) | (6) | (7) | (8) | (9) | (10) |
|---|---|---|---|---|---|---|---|---|---|---|
| 119 | LSBGF304-013 | 045010.5−394745 | 2052 | 22 | 10 | 12.27 | 2.03 | 0.66 | + | |
| 120 | ESO552-016 | 045228.5−191734 | 2918 | 5 | 8 | 11.45 | 2.32 | 0.58 | | F |
| 121 | DDO229 | 045253.0−251445 | 1214 | 9 | 8 | 11.11 | 1.85 | 1.40 | + | IR |
| 122 | NGC1705 | 045413.5−532140 | 401 | 29 | 9 | 10.52 | 2.20 | 1.18 | + | IR |
| 123 | ESO552-031 | 045803.2−190727 | 1527 | 5 | 7 | 13.67 | 1.81 | 0.46 | + | |
| 124 | CGCG445-001 | 045902.7+122403 | 3412 | 5 | 2 | 11.24 | 2.66 | 0.36 | | IR |
| 125 | ESO252-007 | 050301.3−431756 | 2742 | 74 | 4 | 11.72 | 2.53 | 0.47 | + | IR |
| 126 | DDO35 | 050324.6+162416 | 1396 | 5 | 8 | 10.82 | 2.57 | 1.25 | + | IR |
| 127 | UGC03247 | 050638.5+084033 | 3328 | 7 | 4 | 11.98 | 2.16 | 0.77 | + | |
| 128 | ESO553-016 | 051105.9−182537 | 3415 | 5 | 8 | 13.36 | 2.05 | 1.01 | | |
| 129 | ESO423-002 | 051516.3−303135 | 1285 | 7 | 6 | 10.28 | 1.85 | 1.25 | + | IR,F |
| 130 | UGC03288 | 051930.9+040746 | 2983 | 9 | 8 | 13.89 | 3.02 | 0.60 | | |
| 131 | MCG+12-06-004 | 052426.4+711134 | 3467 | 10 | 6 | 12.88 | 2.39 | 0.24 | | |
| 132 | ESO253-002 | 052436.2−460241 | 3366 | 9 | 8 | 13.03 | 1.70 | 0.62 | + | |
| 133 | UGC03303 | 052459.5+043018 | 446 | 5 | 8 | 13.09 | 2.10 | 1.44 | + | |
| 134 | ESO553-046 | 052705.7−204041 | 372 | 10 | 9 | 12.19 | 2.47 | 0.43 | + | IR |
| 135 | MCG-03-14-017 | 052814.1−160728 | 2017 | 5 | 6 | 11.15 | 1.73 | 1.13 | + | IR |
| 136 | ESO554-002 | 052907.6−195602 | 2756 | 39 | 6 | 12.47 | 1.64 | 0.78 | + | IR |
| 137 | MCG-02-15-001 | 053139.9−102332 | 2535 | 36 | 4 | 10.08 | 1.90 | 0.41 | + | IR,F |
| 138 | ESO487-030 | 053718.7−262552 | 1290 | 8 | 7 | 10.94 | 1.85 | 1.11 | + | |
| 139 | ESO306-013 | 053858.3−414414 | 788 | 75 | 9 | 11.04 | 1.67 | 0.65 | + | IR |
| 140 | [KKS2000]54 | 054155.2−123336 | 2090 | 13 | 10 | 11.94 | 1.73 | 0.55 | + | |
| 141 | IC2147 | 054328.1−302942 | 1093 | 5 | 7 | 10.76 | 1.92 | 1.20 | + | IR |
| 142 | ESO253-019 | 054401.9−453017 | 821 | 75 | 10 | 14.58 | 2.21 | 0.3 : | | |
| 143 | ESO488-017 | 054729.0−233434 | 837 | 11 | 4 | 11.67 | 2.43 | 0.63 | + | IR |
| 144 | ESO555-002 | 055026.7−194333 | 2191 | 7 | 4 | 10.00 | 1.66 | 1.34 | + | IR,F |
| 145 | ESO120-016 | 055135.3−590244 | 3415 | 10 | 3 | 9.81 | 2.18 | 0.98 | + | IR,F |
| 146 | 2MASX... | 055302.3−114420 | 3039 | 8 | 3 | 11.68 | 2.52 | 0.88 | + | IR |
| 147 | HIPASSJ0554-35 | 055354.2−355729 | 2720 | 9 | 9 | 13.75 | 1.60 | 0.62 | + | |
| 148 | ESO488-044 | 055550.8−224825 | 3104 | 10 | 4 | 13.37 | 1.83 | 0.56 | + | |
| 149 | ESO425-001 | 060010.3−314714 | 1128 | 9 | 9 | 12.02 | 1.71 | 0.99 | | |
| 150 | RFGC1042 | 060143.4−345642 | 1065 | 10 | 8 | 14.41 | 2.12 | 0.76 | | F |
| 151 | UGC03394 | 060449.6+560957 | 1945 | 5 | 7 | 12.55 | 2.24 | 0.78 | + | |
| 152 | UGC03403 | 061032.9+712245 | 1444 | 5 | 5 | 9.96 | 1.67 | 1.03 | + | IR,F |
| 153 | UGC03409 | 061052.6+643403 | 1516 | 5 | 10 | 14.19 | 2.01 | 1.01 | + | |
| 154 | ESO121-020 | 061554.2−574332 | 307 | 5 | 10 | 13.75 | 2.04 | 0.90 | | |
| 155 | ESO206-016 | 063109.7−522507 | 918 | 5 | 10 | 13.41 | 1.98 | 0.76 | + | |
| 156 | UGC03485 | 063456.3+655008 | 1438 | 7 | 7 | 14.80 | 1.91 | | | F |
| 157 | ESO206-017 | 063819.6−515710 | 753 | 10 | 6 | 13.55 | 1.67 | 0.68 | + | F |
| 158 | ESO308-022 | 063932.7−404315 | 557 | 5 | 10 | 13.27 | 2.10 | 0.51 | + | |
| 159 | [HS98] 011 | 064332.3+634226 | 3042 | 15 | 10 | 14.37 | 2.17 | 0.38 | | |
| 160 | ESO255-019 | 064548.2−473152 | 785 | 5 | 8 | 11.59 | 1.72 | 1.27 | + | |
| 161 | ESO309-005 | 065302.9−391613 | 1634 | 10 | 3 | 11.19 | 2.25 | 0.81 | + | IR |
| 162 | ARGO | 070518.8−583113 | 279 | 9 | 8 | 11.33 | 1.90 | 1.54 | + | |
| 163 | UGC03672 | 070627.6+301919 | 964 | 9 | 10 | 13.74 | 2.10 | 1.48 | | pec |
| 164 | ESO088-004 | 071006.5−631544 | 2037 | 52 | 1 | 10.85 | 1.70 | 0.3 : | + | IR |
| 165 | ESO035-001 | 071042.4−733037 | 2819 | 28 | 4 | 10.47 | 2.34 | 0.47 | + | IR |
| 166 | UGC03761 | 071504.3+380843 | 3359 | 5 | 7 | 12.71 | 2.18 | 0.58 | + | F |
| 167 | ESO162-015 | 071523.7−550435 | 2537 | 10 | 5 | 12.45 | 1.87 | 0.89 | | |
| 168 | UGC03748 | 071527.3+652629 | 2620 | 12 | 7 | 13.61 | 2.12 | 0.88 | + | |
| 169 | CGCG309-028 | 071804.4+682034 | 2796 | 27 | 0 | 10.87 | 2.12 | | + | |
| 170 | UGC03826 | 072427.9+614138 | 1854 | 5 | 7 | 13.00 | 1.84 | 1.42 | + | IR,KIG |
| 171 | UGC03845 | 072642.7+470538 | 3083 | 5 | 4 | 10.72 | 1.87 | 0.71 | | IR |
| 172 | UGC03876 | 072917.5+275358 | 811 | 5 | 6 | 12.89 | 2.00 | 1.06 | + | IR,F,KIG |
| 173 | SDSS... | 073058.9+410960 | 892 | 5 | 9 | 13.76 | 1.89 | | + | |
| 174 | ESO059-001 | 073118.2−681117 | 246 | 9 | 8 | 11.20 | 1.68 | 1.24 | + | |
| 175 | ESO035-012 | 074001.1−762619 | 1190 | 9 | 10 | 12.44 | 1.71 | 0.72 | + | |
| 176 | KUG0738+493 | 074232.4+491130 | 3015 | 14 | 7 | 11.96 | 1.97 | | + | |
| 177 | UGC04115 | 075701.8+142327 | 213 | 5 | 10 | 12.16 | 1.78 | 1.19 | + | |
| 178 | UGC04117 | 075726.0+355621 | 756 | 5 | 8 | 12.40 | 1.82 | 0.68 | + | |
| 179 | NGC2504 | 075952.3+053630 | 2419 | 75 | 4 | 11.42 | 2.29 | 0.76 | + | IR,pec |

| (1) | (2) | (3) | (4) | (5) | (6) | (7) | (8) | (9) | (10) |
|---|---|---|---|---|---|---|---|---|---|
| 180 | UGC04176 | 080242.8+404043 | 3092 | 41 | 7 | 12.65 | 2.29 | 0.70 + | pec |
| 181 | UGC04204 | 080524.9+555700 | 3063 | 8 | 8 | 14.17 | 2.32 | 0.43 + | |
| 182 | CGCG148-111 | 080547.7+301401 | 2295 | 36 | 6 | 13.16 | 2.45 | + | F |
| 183 | UGC04247 | 080903.6+164039 | 2717 | 8 | 7 | 12.93 | 2.60 | 1.22 | F |
| 184 | UGC04254 | 080924.0+003634 | 1610 | 5 | 5 | 10.62 | 1.81 | 0.75 + | IR |
| 185 | UGC04258 | 081047.8+465444 | 3172 | 8 | 6 | 12.08 | 1.62 | 0.67 + | F |
| 186 | UGC04267 | 081256.4+545808 | 2694 | 75 | 6 | 11.73 | 2.26 | + | IR |
| 187 | UGC04307 | 081801.4+475109 | 3164 | 5 | 6 | 12.46 | 1.62 | | |
| 188 | NGC2574 | 082048.2−085506 | 2635 | 5 | 4 | 9.79 | 1.67 | 0.95 | |
| 189 | SDSS... | 082712.8+265127 | 1779 | 5 | 9 | 14.86 | 2.06 | + | |
| 190 | CGCG032-035 | 083121.6+070000 | 1702 | 34 | 8 | 12.85 | 1.72 | + | |
| 191 | UGC04537 | 084244.9+354528 | 2883 | 5 | 6 | 13.49 | 2.53 | 0.70 + | F |
| 192 | MCG-01-23-002 | 084900.4−074947 | 2667 | 7 | 4 | 10.55 | 2.30 | 0.65 + | IR,F |
| 193 | ESO060-015 | 085003.7−700736 | 3284 | 10 | 5 | 11.45 | 1.75 | 0.77 + | IR,F |
| 194 | LSBCD563-06 | 085222.8+210050 | 3456 | 5 | 10 | 15.32 | 1.66 | 0.3 : + | |
| 195 | UGC04684 | 085640.7+002230 | 2316 | 7 | 7 | 12.23 | 2.00 | 0.85 + | IR,KIG |
| 196 | NGC2722 | 085846.2−034236 | 2535 | 8 | 3 | 10.41 | 1.73 | 0.86 + | IR |
| 197 | UGC04722 | 090023.5+253641 | 1705 | 7 | 7 | 12.14 | 2.15 | 1.29 + | IR,F,pec,KIG |
| 198 | KUG0857+479 | 090058.5+474743 | 3175 | 48 | 4 | 11.19 | 1.63 | + | |
| 199 | NGC2731 | 090208.4+081806 | 2385 | 23 | 6 | 10.47 | 2.07 | 0.42 + | IR |
| 200 | UGC04711 | 090322.8+784505 | 3480 | 40 | 3 | 12.69 | 1.93 | + | |
| 201 | KKH46 | 090836.5+051727 | 409 | 5 | 10 | 15.71 | 1.68 | 0.42 | |
| 202 | LSBC D634-03 | 090853.5+143455 | 181 | 12 | 10 | 14.98 | 1.69 | −0.60 + | |
| 203 | CGCG121-027 | 090934.4+251323 | 2200 | 37 | 9 | 12.75 | 2.34 | 0.3 : + | |
| 204 | ESO006-005 | 090947.4−833130 | 1771 | 20 | 7 | 13.87 | 1.61 | 0.59 + | |
| 205 | ESO018-015 | 091011.5−791404 | 1431 | 10 | 6 | 11.78 | 1.65 | 1.07 + | IR |
| 206 | CGCG006-011 | 091125.3−025257 | 3371 | 10 | 5 | 13.03 | 1.71 | 0.60 + | IR |
| 207 | SDSS... | 091126.7+455226 | 3415 | 20 | 6 | 13.77 | 1.99 | + | |
| 208 | IC2450 | 091705.3+252545 | 1607 | 32 | 3 | 10.92 | 1.90 | 0.3 : | IR,Mrk |
| 209 | UGC04925 | 091819.8+174512 | 2881 | 5 | 7 | 12.18 | 2.00 | 1.02 + | F |
| 210 | UGC04922 | 091836.5+475221 | 2022 | 5 | 8 | 11.23 | 1.70 | 1.36 + | |
| 211 | FGC0878 | 092050.4−034659 | 3262 | 12 | 8 | 14.49 | 2.17 | 0.3 : + | F |
| 212 | CGCG062-024 | 092059.6+110333 | 1128 | 15 | 9 | 12.99 | 1.78 | 0.3 : | |
| 213 | KUG0917+461 | 092110.8+455316 | 1884 | 41 | 4 | 11.58 | 1.70 | + | IR,KIG |
| 214 | UGC04970 | 092145.6+393129 | 2417 | 47 | 5 | 12.03 | 2.03 | 0.58 + | F |
| 215 | CGCG151-073 | 092310.9+264905 | 2375 | 7 | 9 | 13.23 | 1.68 | 0.48 + | |
| 216 | UGC05023 | 092601.2+192301 | 2404 | 10 | 9 | 11.73 | 2.26 | 0.37 | IR,Mrk |
| 217 | UGC05078 | 093145.8+034343 | 3021 | 8 | 7 | 13.58 | 1.93 | 1.18 + | F |
| 218 | UGC05135 | 093836.2+431037 | 1717 | 67 | 6 | 13.17 | 2.14 | | |
| 219 | UGC05114 | 094003.2+820617 | 1811 | 11 | 9 | 13.27 | 1.84 | 0.49 + | pec |
| 220 | 6dF... | 094208.4−233544 | 3043 | 74 | 11 | 11.05 | 1.66 | 0.3 : + | |
| 221 | SBS0945+594 | 094841.6+591539 | 2313 | 5 | 4 | 11.87 | 1.86 | + | IR,Mrk |
| 222 | MRK1426 | 094918.4+483350 | 1906 | 26 | 4 | 13.67 | 2.14 | 0.19 | Mrk |
| 223 | SDSS... | 095058.8+104805 | 3017 | 5 | 5 | 13.24 | 1.81 | 1.00 | |
| 224 | UGC05299 | 095241.3−001103 | 2705 | 8 | 8 | 12.99 | 1.98 | 0.78 + | |
| 225 | 6dF... | 095536.6−165756 | 2759 | 74 | 8 | 12.72 | 1.92 | 0.3 : | |
| 226 | UGC05309 | 095711.4+804435 | 3252 | 8 | 8 | 14.41 | 1.98 | 0.52 | |
| 227 | KUG0956+420[1] | 095930.0+414601 | 1678 | 36 | 9 | 14.54 | 2.13 | | |
| 228 | KUG0956+457 | 100005.2+453111 | 1708 | 5 | 9 | 14.60 | 1.81 | | |
| 229 | ESO567-012 | 100526.0−174757 | 2780 | 75 | 4 | 11.05 | 1.64 | 0.3 : + | |
| 230 | KUG1003+466 | 100646.7+462304 | 2417 | 38 | 7 | 14.37 | 1.96 | | |
| 231 | UGC05467 | 100812.9+184225 | 2768 | 26 | −2 | 10.83 | 2.10 | 0.64 + | IR |
| 232 | HIPASSJ1008-33 | 100828.7−330837 | 1337 | 9 | 10 | 14.30 | 1.81 | 0.50 | |
| 233 | SBS1006+578 | 100935.4+573401 | 1593 | 5 | 9 | 14.33 | 1.70 | 0.51 + | Mrk |
| 234 | NGC3139 | 101005.2−114642 | 1157 | 10 | 1 | 10.25 | 1.98 | 0.3 : | |
| 235 | UGC05493 | 101117.9+002633 | 3436 | 5 | 5 | 11.08 | 2.30 | 0.78 + | IR |
| 236 | UGC05526 | 101428.6+155411 | 2928 | 40 | 5 | 12.72 | 1.72 | 0.3 : | |

[1] Probably this galaxy makes a pair with KUG0956+419 having $V_{LG} = 1733$ km/s

| (1) | (2) | (3) | (4) | (5) | (6) | (7) | (8) | (9) | (10) |
|---|---|---|---|---|---|---|---|---|---|
| 237 | SDSS... | 101456.7+604557 | 3331 | 95 | 8 | 13.44 | 1.99 | | + | |
| 238 | UGCA208 | 101628.2+451918 | 1671 | 5 | 9 | 12.65 | 1.88 | 0.65 | | Mrk |
| 239 | IC2563 | 101851.9−323548 | 1044 | 5 | 7 | 12.64 | 1.88 | 0.54 | + | |
| 240 | ESO567-048 | 101941.6−174460 | 631 | 5 | 8 | 10.28 | 1.90 | 1.11 | | |
| 241 | MRK0630 | 102310.4+175746 | 3425 | 45 | 9 | 11.67 | 1.93 | 0.5 : | | IR,Mrk |
| 242 | KUG1028+412 | 103118.4+410226 | 2568 | 5 | 9 | 15.23 | 2.12 | | | |
| 243 | PGC031148 | 103238.2−172534 | 2377 | 5 | 8 | 13.09 | 1.78 | 0.64 | + | pec |
| 244 | MRK1434 | 103410.1+580349 | 2247 | 75 | 9 | 15.24 | 1.62 | | | Mrk |
| 245 | UGC05744 | 103504.8+463341 | 3365 | 30 | 1 | 11.04 | 1.74 | | | IR,Mrk |
| 246 | UGC05825 | 104211.1+234448 | 3390 | 8 | 4 | 10.78 | 2.15 | 0.90 | + | |
| 247 | DDO87 | 104936.5+653150 | 467 | 5 | 10 | 12.83 | 1.76 | 1.21 | | |
| 248 | CGCG038-048 | 105512.6+055145 | 3320 | 18 | 3 | 12.41 | 2.05 | 0.5 : | + | |
| 249 | ESO437-071 | 105519.3−302804 | 1875 | 5 | 6 | 10.90 | 1.71 | 1.00 | + | |
| 250 | MCG-02-28-031 | 105938.2−153135 | 2785 | 6 | 6 | 11.45 | 2.12 | 1.01 | + | F |
| 251 | UGC06138 | 110439.7+274326 | 2506 | 5 | 5 | 11.00 | 1.83 | 0.94 | | |
| 252 | 2MASX... | 110443.6−290633 | 2103 | 74 | 6 | 12.49 | 1.76 | 0.5 : | + | F |
| 253 | ESO438-002 | 110542.0−312737 | 3426 | 5 | 6 | 13.25 | 2.00 | 0.5 : | + | F |
| 254 | KKSG23 | 110611.5−142432 | 789 | 5 | 10 | 13.22 | 1.61 | 1.04 | + | |
| 255 | 2MASX... | 110703.8−173622 | 749 | 29 | 8 | 13.09 | 1.69 | 0.74 | | |
| 256 | CGCG364-019 | 110734.3+825114 | 1876 | 26 | 0 | 13.06 | 1.96 | | + | |
| 257 | KUG1107+403 | 111025.2+400311 | 2932 | 29 | 5 | 13.42 | 1.89 | | + | |
| 258 | PGC034171 | 111329.0−061525 | 2312 | 5 | 6 | 11.17 | 1.77 | 0.29 | | |
| 259 | KK100 | 111359.3+111944 | 2822 | 5 | 7 | 14.68 | 1.70 | 0.3 : | + | |
| 260 | ESO265-018 | 111436.1−431610 | 2608 | 39 | 10 | 13.66 | 1.80 | 0.65 | | pec |
| 261 | UGC06383 | 112202.6+424909 | 3177 | 8 | 6 | 13.46 | 2.12 | 0.72 | | IR,F,KIG |
| 262 | MCG-03-29-006 | 112232.9−173412 | 3399 | 42 | 4 | 13.12 | 1.68 | 0.3 : | + | IR |
| 263 | UGC06517 | 113202.4+364153 | 2472 | 9 | 4 | 11.10 | 1.85 | 0.82 | + | IR |
| 264 | 2MASX... | 114234.8−165210 | 2226 | 40 | −3 | 10.90 | 1.62 | 0.5 : | | |
| 265 | SDSS... | 114805.4+005929 | 2867 | 66 | 8 | 14.93 | 1.69 | 0.5 : | | |
| 266 | ESO504-016 | 114834.4−255710 | 2890 | 34 | 1 | 12.46 | 1.72 | 0.3 : | + | |
| 267 | 6dF... | 115042.6−101312 | 2131 | 42 | 8 | 12.74 | 1.63 | 0.43 | + | |
| 268 | UGC06890 | 115511.7+002915 | 3038 | 5 | 8 | 13.84 | 1.72 | 0.82 | + | |
| 269 | ESO020-003 | 115514.0−784436 | 2733 | 9 | 8 | 14.33 | 2.05 | 0.56 | + | |
| 270 | NGC4025 | 115910.2+374737 | 3217 | 7 | 6 | 12.33 | 2.07 | 0.87 | | F |
| 271 | 6dF... | 120227.5−190602 | 2268 | 74 | 9 | 13.38 | 1.87 | 0.3 : | + | |
| 272 | ESO505-025 | 121045.1−264215 | 1625 | 74 | 9 | 13.36 | 1.75 | 0.48 | + | |
| 273 | AM1213-220 | 121557.8−222530 | 2141 | 5 | 10 | 14.36 | 1.89 | 0.62 | + | pec |
| 274 | SDSS... | 122703.0+413423 | 2104 | 5 | 8 | 14.07 | 1.75 | | | |
| 275 | MCG-02-32-012 | 122758.8−133130 | 3249 | 37 | −2 | 11.75 | 1.95 | 0.3 : | + | |
| 276 | NGC4529 | 123251.6+201101 | 2468 | 5 | 5 | 12.93 | 1.72 | 0.86 | | F |
| 277 | ESO442-013 | 123713.5−282934 | 1269 | 6 | 6 | 11.90 | 1.63 | 1.54 | + | |
| 278 | UGCA291 | 123837.3+555533 | 3364 | 58 | 9 | 13.69 | 2.30 | | + | F,Mrk |
| 279 | SDSS... | 124417.4+594308 | 3054 | 5 | 9 | 14.72 | 1.65 | | + | |
| 280 | IC3740 | 124530.6+204857 | 2602 | 65 | 4 | 12.81 | 2.16 | 0.7 : | | |
| 281 | SBS1245+542 | 124809.9+540127 | 3437 | 69 | 8 | 11.96 | 2.30 | | + | IR,Mrk |
| 282 | SDSS... | 125446.3+153530 | 2564 | 11 | 9 | 14.29 | 1.76 | 0.3 : | | |
| 283 | MCG+09-21-092 | 130255.7+554140 | 1472 | 14 | 9 | 13.94 | 1.84 | 0.55 | | F |
| 284 | MCG+08-24-030 | 130321.3+481951 | 2595 | 56 | 5 | 12.83 | 1.70 | | + | |
| 285 | UGC08166 | 130352.5+105821 | 2853 | 5 | 6 | 12.68 | 1.92 | 0.94 | + | F,KIG |
| 286 | UGC08224 | 130821.6+392740 | 3328 | 75 | 5 | 12.69 | 1.66 | 0.73 | + | F |
| 287 | CGCG101-017 | 131556.3+174538 | 1178 | 37 | 9 | 14.48 | 1.93 | 0.3 : | + | KIG |
| 288 | DDO171 | 131841.2−082647 | 1150 | 5 | 8 | 11.38 | 1.83 | 0.70 | + | |
| 289 | NGC5089 | 131939.3+301523 | 2138 | 11 | 3 | 10.86 | 1.89 | 0.92 | + | IR |
| 290 | NGC5116 | 132255.6+265850 | 2887 | 5 | 5 | 9.90 | 2.01 | 0.88 | | IR |
| 291 | UGC08509 | 132818.9+673753 | 1162 | 5 | 9 | 13.89 | 1.94 | | + | |
| 292 | SDSS... | 133047.8+395446 | 1290 | 5 | 9 | 15.53 | 1.65 | | | |
| 293 | UGCA362 | 133305.9+685138 | 1676 | 75 | 9 | 13.53 | 1.78 | | + | IR,Mrk |
| 294 | UGC08578 | 133535.6+291301 | 872 | 10 | 9 | 13.31 | 2.28 | 0.54 | | Mrk |
| 295 | DDO180 | 133810.3−094805 | 1153 | 9 | 8 | 9.99 | 1.81 | 1.15 | + | IR |
| 296 | UGC08647 | 133948.1+311725 | 776 | 5 | 10 | 14.08 | 2.18 | 1.01 | | |
| 297 | MCG-01-35-010 | 134537.0−055923 | 1333 | 11 | 8 | 12.90 | 1.75 | 1.66 | | IR |

| (1) | (2) | (3) | (4) | (5) | (6) | (7) | (8) | (9) | (10) |
|---|---|---|---|---|---|---|---|---|---|
| 298 | KK220 | 134736.5+331222 | 812 | 5 | 10 14.98 | 2.18 | −0.08 | + | |
| 299 | UGC08737 | 134816.7+680506 | 1959 | 5 | 4 9.69 | 1.65 | 0.54 | + | IR,F |
| 300 | ESO577-038 | 134825.7−185220 | 1719 | 5 | 7 13.49 | 2.70 | 1.25 | − | F |
| 301 | KK224 | 134857.3+433601 | 1244 | 5 | 10 14.79 | 1.71 | 0.26 | | |
| 302 | HIPASSJ1349-12 | 134910.0−124535 | 1246 | 9 | 8 12.99 | 1.81 | 0.83 | + | pec |
| 303 | CGCG102-075 | 135305.4+155040 | 3019 | 41 | 1 11.10 | 2.65 | 0.5 : | | |
| 304 | DDO184 | 135524.9+174743 | 939 | 5 | 8 10.89 | 1.81 | 1.25 | − | pec |
| 305 | UGC08894 | 135747.6+632351 | 1938 | 11 | 8 13.82 | 1.73 | 0.86 | + | |
| 306 | ESO008-004 | 135748.0−830449 | 2208 | 5 | 6 11.72 | 1.64 | 1.16 | − | IR |
| 307 | MCG+10-20-057 | 135842.9+615146 | 1658 | 9 | 9 13.87 | 1.79 | | + | |
| 308 | CGCG018-021 | 140043.0−003020 | 3361 | 21 | 3 13.71 | 1.93 | 0.06 | + | |
| 309 | NGC5470 | 140632.0+060146 | 962 | 5 | 4 9.96 | 1.75 | 0.96 | + | IR,F |
| 310 | UGC09024 | 140640.6+220412 | 2331 | 5 | 2 13.49 | 1.61 | 0.92 | + | pec |
| 311 | ESO511-008 | 141128.6−261214 | 2440 | 35 | 2 11.09 | 1.69 | 0.38 | + | IR |
| 312 | NGC5510 | 141337.2−175902 | 1309 | 19 | 8 11.04 | 1.67 | 1.16 | + | IR |
| 313 | NGC5523 | 141452.3+251903 | 1073 | 5 | 6 9.74 | 1.96 | 1.46 | + | IR,F,KIG |
| 314 | PGC051218 | 141960.0−101150 | 3286 | 29 | 7 13.37 | 1.74 | 0.65 | + | |
| 315 | UGC09193 | 142113.8+364434 | 770 | 5 | 10 14.62 | 2.23 | 0.49 | + | |
| 316 | ESO446-055 | 142129.4−275602 | 2416 | 6 | 4 12.89 | 2.07 | 0.59 | + | |
| 317 | DDO189 | 142232.2+452302 | 803 | 5 | 8 13.13 | 1.63 | 1.41 | + | |
| 318 | HIPASSJ1424-16B | 142429.0−165858 | 1352 | 9 | 8 12.01 | 1.67 | 1.10 | | |
| 319 | DDO190 | 142443.4+443133 | 263 | 6 | 10 11.00 | 2.49 | 1.39 | + | |
| 320 | KKSG46 | 142826.6−085516 | 1439 | 9 | 10 14.30 | 1.60 | 0.74 | + | |
| 321 | UGC09320 | 142958.4+365224 | 859 | 6 | 8 13.88 | 2.21 | 0.69 | + | |
| 322 | ESO385-032 | 143011.9−365752 | 2646 | 5 | 5 9.71 | 1.74 | 1.04 | + | IR |
| 323 | 6dF... | 143047.8−091327 | 2486 | 74 | 8 12.82 | 2.06 | 0.5 : | | |
| 324 | SDSS... | 143208.7+383122 | 1492 | 71 | 9 14.75 | 2.28 | | | |
| 325 | LSBCD512-02 | 143320.1+265950 | 883 | 5 | 10 13.08 | 2.70 | 0.38 | + | |
| 326 | KKSG47 | 143525.0−170947 | 1448 | 5 | 10 15.03 | 2.25 | 0.96 | + | |
| 327 | 2MASX... | 143839.3−105407 | 3338 | 74 | 3 12.03 | 2.05 | 0.5 : | + | IR |
| 328 | MRK0475 | 143905.5+364822 | 655 | 30 | 13 13.95 | 2.13 | −0.82 | + | Mrk |
| 329 | ESO386-013 | 144304.3−332926 | 1205 | 9 | 10 12.41 | 2.64 | 0.67 | + | |
| 330 | SDSS... | 144310.9+382045 | 2875 | 30 | 6 13.40 | 2.10 | | + | |
| 331 | UGC09497 | 144412.8+423744 | 884 | 42 | 7 13.16 | 1.74 | | + | |
| 332 | UGC09519 | 144621.1+342214 | 1782 | 26 | 0 10.23 | 1.62 | | + | IR |
| 333 | SDSS... | 144827.7+385233 | 1938 | 37 | 9 14.58 | 2.55 | | | |
| 334 | MCG-01-38-003 | 144848.0−034259 | 868 | 9 | 9 12.22 | 1.90 | 1.39 | + | |
| 335 | UGC09540 | 144852.0+344242 | 895 | 5 | 14 14.80 | 2.30 | 0.68 | + | |
| 336 | ESO386-029 | 145323.3−360511 | 1392 | 74 | 8 12.12 | 2.64 | 0.38 | + | |
| 337 | UGC09588 | 145411.9+301232 | 2939 | 16 | 9 12.47 | 1.96 | | + | IR |
| 338 | 2MASX... | 145739.4+263953 | 1477 | 75 | 3 13.63 | 2.30 | | + | pec |
| 339 | NGC5832 | 145745.7+714056 | 657 | 5 | 8 10.28 | 1.85 | 1.44 | + | IR,KIG |
| 340 | ESO581-012 | 150004.9−223926 | 3350 | 9 | 7 13.79 | 2.19 | 0.92 | | IR,F,pec |
| 341 | UGC09676 | 150330.6+274929 | 2962 | 5 | 7 12.14 | 2.53 | | | |
| 342 | UGC09730 | 150403.0+773804 | 2353 | 11 | 6 12.78 | 2.38 | 0.78 | + | IR,KIG |
| 343 | PGC1067957 | 150704.8−034202 | 1638 | 74 | 9 13.36 | 2.05 | 0.48 | + | |
| 344 | PGC054003 | 150737.3−175450 | 2984 | 5 | 8 14.32 | 1.69 | 1.03 | + | |
| 345 | UGC09739 | 150855.4+254402 | 1441 | 5 | 7 12.56 | 2.30 | 0.73 | + | KIG |
| 346 | UGC09764 | 151038.2+645354 | 2444 | 11 | 7 13.18 | 1.97 | 1.19 | + | |
| 347 | ESO513-022 | 151200.7−240236 | 3469 | 74 | 4 11.71 | 1.77 | 0.92 | | IR,pec |
| 348 | FGC1874 | 151532.1+493714 | 2668 | 14 | 7 14.47 | 2.18 | 0.10 | | F |
| 349 | CGCG221-048 | 151701.1+394144 | 978 | 39 | 9 13.66 | 2.99 | | | |
| 350 | CGCG106-029 | 151754.3+161841 | 766 | 5 | 10 13.22 | 2.88 | 0.11 | + | |
| 351 | SDSS... | 151832.1+310940 | 1626 | 61 | 10 14.57 | 2.44 | | | |
| 352 | UGC09814 | 151925.1+110315 | 3256 | 5 | 8 12.75 | 2.78 | 0.51 | + | IR |
| 353 | IC4538 | 152111.6−233930 | 2755 | 18 | 5 9.45 | 1.83 | 0.80 | + | IR |
| 354 | ESO449-009 | 152311.9−322419 | 3243 | 74 | 2 10.82 | 2.81 | 0.72 | + | |
| 355 | UGC09880 | 153044.4+471908 | 2730 | 5 | 8 13.83 | 2.40 | 0.69 | | |
| 356 | UGC09875 | 153047.5+230357 | 2072 | 5 | 8 14.18 | 1.95 | 0.68 | + | |
| 357 | UGC09893 | 153257.3+462710 | 817 | 5 | 10 12.61 | 2.75 | 1.02 | + | KIG |
| 358 | KKR21 | 153700.6+204742 | 1798 | 5 | 8 13.22 | 2.51 | 0.75 | + | |

| (1) | (2) | (3) | (4) | (5) | (6) | (7) | (8) | (9) | (10) |
|---|---|---|---|---|---|---|---|---|---|
| 359 | 6dF... | 153729.3−285308 | 2618 | 74 | 8 | 12.65 | 1.82 | 0.7 : | + | |
| 360 | NGC5964 | 153736.3+055824 | 1468 | 5 | 7 | 11.79 | 1.67 | 1.46 | + | IR,KIG |
| 361 | SDSS... | 153928.7+005637 | 3006 | 22 | 7 | 13.82 | 2.27 | 0.5 : | + | |
| 362 | UGC10023 | 154609.7+065354 | 1440 | 5 | 8 | 12.01 | 1.67 | 0.70 | + | |
| 363 | UGC10025 | 154624.4+025039 | 1538 | 5 | 8 | 13.90 | 2.26 | 0.71 | | F |
| 364 | UGC10058 | 155024.2+255521 | 2257 | 5 | 8 | 14.16 | 2.08 | 0.72 | | |
| 365 | UGC10125 | 155206.6+835136 | 1867 | 5 | 8 | 13.77 | 2.08 | 1.08 | + | |
| 366 | UGC10119 | 155438.5+791025 | 2441 | 19 | 4 | 13.18 | 2.41 | | | |
| 367 | HIPASSJ1558-10 | 155820.0−103207 | 906 | 9 | 8 | 12.06 | 1.84 | 1.07 | + | |
| 368 | CGCG319-040 | 160227.9+642112 | 1813 | 20 | 6 | 13.16 | 2.42 | | | IR |
| 369 | UGC10175 | 160437.7+304253 | 952 | 6 | 8 | 12.46 | 2.48 | 0.66 | | |
| 370 | 2MASX... | 160511.0+462331 | 3155 | 74 | 5 | 13.57 | 2.50 | | + | |
| 371 | KKSG48 | 160540.9−043419 | 1616 | 5 | 10 | 12.64 | 1.60 | 0.86 | + | |
| 372 | 2MASX... | 160705.2−112552 | 1786 | 74 | 4 | 11.43 | 2.59 | 0.5 : | + | |
| 373 | UGC10247 | 160926.0+600517 | 3210 | 5 | 8 | 12.77 | 1.71 | 0.76 | + | |
| 374 | MCG-01-41-006 | 160936.8−043713 | 847 | 9 | 10 | 11.92 | 2.38 | 0.84 | | |
| 375 | SHOC 529 | 161111.5+482004 | 3011 | 16 | 9 | 13.15 | 2.20 | | | |
| 376 | UGC10281 | 161320.6+171134 | 1182 | 5 | 10 | 15.39 | 3.00 | 0.67 | + | |
| 377 | HIPASSJ1615-17 | 161545.3−175030 | 2330 | 9 | 8 | 11.96 | 3.32 | 0.90 | + | |
| 378 | DDO204 | 161618.3+470247 | 908 | 6 | 8 | 11.03 | 2.61 | 1.17 | + | IR |
| 379 | KKR26 | 161644.6+160509 | 2347 | 5 | 10 | 14.56 | 1.79 | 0.36 | | |
| 380 | MCG-02-41-001 | 161715.8−114355 | 960 | 5 | 3 | 9.80 | 1.84 | 1.60 | + | IR |
| 381 | 2MASX... | 162047.4+470354 | 3218 | 35 | 9 | 14.05 | 2.34 | | + | |
| 382 | LSBCD584-05 | 162123.8+205156 | 3220 | 5 | 8 | 14.01 | 1.76 | 1.27 | | |
| 383 | UGC10383 | 162442.8+643044 | 2990 | 5 | 8 | 12.31 | 2.30 | 0.82 | + | |
| 384 | UGC10419 | 163006.1+274158 | 2762 | 5 | 8 | 14.34 | 2.38 | 0.62 | | |
| 385 | UGC10437 | 163107.6+432055 | 2795 | 13 | 6 | 12.71 | 3.08 | 1.10 | | KIG |
| 386 | SDSS... | 163424.7+245741 | 1131 | 40 | 9 | 15.51 | 1.85 | | + | |
| 387 | UGCA412 | 163521.1+521253 | 2865 | 33 | 9 | 13.05 | 2.39 | −0.35 | + | IR,Mrk |
| 388 | PGC165693 | 164302.7−204009 | 1174 | 5 | 7 | 10.33 | 2.80 | 0.47 | + | IR |
| 389 | 2MASX... | 164819.4−103139 | 1595 | 9 | 8 | 11.54 | 2.56 | 1.07 | + | |
| 390 | UGC10589 | 165025.0+555027 | 2351 | 11 | 6 | 12.25 | 3.00 | | + | |
| 391 | DDO206 | 165421.5+530647 | 1317 | 5 | 8 | 12.59 | 1.88 | 0.95 | + | IR |
| 392 | KKR30 | 165638.5+075955 | 1584 | 6 | 8 | 13.87 | 2.01 | 0.67 | + | |
| 393 | NGC6283 | 165926.6+495519 | 1317 | 36 | 4 | 10.83 | 1.88 | | + | IR |
| 394 | HIPASSJ1700-12 | 170006.3−120001 | 1327 | 9 | 8 | 13.46 | 3.07 | 0.76 | | |
| 395 | UGC10721 | 170825.6+253103 | 3091 | 5 | 4 | 10.57 | 2.12 | 0.72 | + | |
| 396 | KKR34 | 171242.1+135428 | 1640 | 5 | 10 | 14.78 | 1.93 | 0.61 | + | |
| 397 | NGC6339 | 171706.5+405042 | 2326 | 6 | 6 | 10.96 | 1.89 | 1.05 | | IR |
| 398 | DDO207 | 171951.0+142401 | 1698 | 5 | 9 | 12.61 | 1.88 | 0.64 | + | KIG |
| 399 | CGCG226-010 | 172410.3+432711 | 2382 | 38 | 5 | 12.65 | 1.92 | | | IR |
| 400 | UGC10854 | 172446.4+581221 | 3052 | 11 | 6 | 12.30 | 2.01 | 0.49 | | |
| 401 | UGC10899 | 173335.8+204604 | 3402 | 5 | 4 | 11.92 | 1.85 | 0.76 | + | |
| 402 | UGC10901 | 173354.1+052834 | 2956 | 8 | 6 | 13.62 | 2.91 | 0.96 | + | |
| 403 | KUG1736+636 | 173637.1+633502 | 1454 | 75 | 9 | 13.00 | 2.51 | | + | IR |
| 404 | IC1265 | 173639.4+420518 | 2397 | 6 | 2 | 10.99 | 1.92 | 0.64 | | IR |
| 405 | NGC6434 | 173648.8+720520 | 2741 | 6 | 4 | 9.96 | 2.07 | 1.00 | + | IR |
| 406 | 6dF... | 173924.3−674906 | 3075 | 9 | 6 | 12.72 | 1.64 | 0.68 | + | |
| 407 | VII Zw 744 | 174137.7+830759 | 2120 | 69 | −1 | 12.02 | 2.37 | | + | KIG |
| 408 | HIPASSJ1745-59 | 174526.7−593128 | 3113 | 9 | 8 | 13.34 | 1.93 | 0.63 | + | |
| 409 | KKR36 | 174616.3+020658 | 3107 | 6 | 9 | 13.59 | 2.45 | 0.59 | + | |
| 410 | UGC10985 | 174804.4+144429 | 1969 | 5 | 8 | 10.71 | 2.11 | 0.68 | | IR,F |
| 411 | MRK1119 | 175236.9+374453 | 3431 | 36 | 0 | 12.04 | 2.47 | | + | IR,Mrk |
| 412 | NGC6542 | 175938.6+612134 | 932 | 35 | 2 | 11.03 | 2.08 | 0.32 | + | IR,F |
| 413 | UGC11109 | 180414.0+464414 | 1820 | 5 | 10 | 14.46 | 2.75 | 0.72 | | |
| 414 | UGC11220 | 182325.5+405643 | 1706 | 5 | 10 | 14.45 | 2.78 | 0.98 | + | |
| 415 | UGC11231 | 182538.3+292232 | 3135 | 75 | 6 | 13.38 | 2.00 | 0.27 | + | F |
| 416 | UGC11295 | 183307.5+752401 | 2632 | 5 | 8 | 13.58 | 2.08 | 1.04 | + | F |
| 417 | NGC6689 | 183450.3+703126 | 758 | 5 | 6 | 10.11 | 1.72 | 1.45 | + | IR,KIG |
| 418 | UGC11382 | 185357.7+743936 | 2679 | 6 | 10 | 13.60 | 2.08 | 0.19 | + | |
| 419 | UGC11400 | 190025.5+790220 | 1991 | 5 | 8 | 14.04 | 2.54 | 0.33 | | |

| (1) | (2) | (3) | (4) | (5) | (6) | (7) | (8) | (9) | (10) |
|---|---|---|---|---|---|---|---|---|---|
| 420 | NGC6762 | 190537.1+635603 | 3205 | 28 | 0 | 10.19 | 2.18 | | + | |
| 421 | IC4819 | 190707.3−592801 | 1728 | 10 | 6 | 11.01 | 1.65 | 1.45 | + | F |
| 422 | NGC6789 | 191641.1+635824 | 136 | 11 | 9 | 12.22 | 1.65 | 0.0 : | + | |
| 423 | NGC6796 | 192131.1+610855 | 2477 | 14 | 4 | 9.56 | 2.32 | 1.38 | + | IR,F |
| 424 | ESO525-005 | 192645.2−272514 | 1995 | 9 | 10 | 13.77 | 2.14 | 0.41 | | |
| 425 | HIPASSJ1926-74 | 192727.1−740458 | 2444 | 9 | 9 | 14.51 | 2.23 | 0.36 | | |
| 426 | 2MASX... | 193159.0−362646 | 2769 | 11 | 5 | 12.99 | 2.31 | 0.11 | | |
| 427 | ESO594-011 | 194042.4−221614 | 1874 | 5 | 5 | 11.97 | 2.46 | 0.23 | + | |
| 428 | UGCA416 | 194822.0−180442 | 1864 | 19 | 8 | 12.34 | 1.93 | 1.02 | + | |
| 429 | 2MASX... | 195015.9−101724 | 2163 | 10 | 4 | 11.63 | 1.86 | 1.06 | + | |
| 430 | UGC11496 | 195302.0+673954 | 2402 | 7 | 8 | 13.19 | 1.72 | 0.80 | + | |
| 431 | KK246 | 200357.4−314054 | 436 | 49 | 10 | 13.56 | 3.22 | 0.92 | + | |
| 432 | IC4968 | 201450.2−644754 | 3348 | 74 | 5 | 11.60 | 1.66 | 0.5 : | | |
| 433 | IC4986 | 201711.6−550211 | 2080 | 16 | 7 | 11.94 | 2.02 | 1.30 | | |
| 434 | FGC2266 | 201729.1−105051 | 2010 | 5 | 6 | 12.72 | 3.20 | 0.99 | + | F |
| 435 | 2MASX... | 201731.5+720726 | 2692 | 22 | −1 | 10.71 | 1.95 | | + | IR |
| 436 | 6dF... | 202031.8−045358 | 1567 | 9 | 8 | 13.66 | 2.15 | 1.06 | + | |
| 437 | UGC11565 | 202802.9+045743 | 3318 | 9 | 6 | 11.71 | 2.68 | 1.03 | + | IR,F |
| 438 | UGC11566 | 202811.9+001718 | 1875 | 35 | 4 | 10.91 | 1.79 | 0.3 : | + | IR,Mrk |
| 439 | AM2024-612 | 202843.2−611819 | 3377 | 74 | 8 | 13.44 | 1.72 | 0.50 | | |
| 440 | KKR54 | 203558.8−011859 | 1929 | 7 | 10 | 14.23 | 1.79 | 0.3 : | + | |
| 441 | 2MASX... | 203857.3−634616 | 1536 | 9 | 9 | 13.60 | 1.94 | 0.57 | + | |
| 442 | IC5028 | 204322.0−653852 | 1498 | 9 | 10 | 12.78 | 1.94 | 0.92 | | |
| 443 | IC5052 | 205201.6−691136 | 447 | 7 | 7 | 8.87 | 2.65 | 2.20 | + | IR |
| 444 | ESO598-017 | 210230.8−192752 | 2546 | 74 | 8 | 13.13 | 2.83 | 0.41 | + | |
| 445 | ESO341-032 | 210333.4−392647 | 2762 | 9 | 9 | 11.53 | 1.64 | 0.98 | + | IR |
| 446 | ESO464-023 | 210656.0−300542 | 2619 | 75 | 8 | 14.29 | 2.20 | 0.3 : | | |
| 447 | ESO187-051 | 210733.1−545702 | 1313 | 32 | 8 | 12.54 | 2.10 | 1.09 | | pec |
| 448 | MCG-01-54-003 | 211005.1−034206 | 2425 | 5 | 5 | 10.94 | 2.26 | 0.56 | + | |
| 449 | ESO107-016 | 211606.2−644901 | 1704 | 48 | 7 | 12.30 | 1.88 | 1.46 | + | F |
| 450 | CGCG471-002 | 211652.9+241215 | 3164 | 29 | 10 | 10.41 | 2.29 | 0.5 : | + | IR |
| 451 | MCG-01-54-016 | 212559.8−034836 | 3127 | 5 | 7 | 13.26 | 2.08 | 0.86 | | F |
| 452 | ESO530-051 | 212744.0−234006 | 1904 | 43 | 8 | 12.89 | 2.31 | 0.54 | + | |
| 453 | NGC7077 | 212959.6+022451 | 1369 | 5 | 9 | 11.36 | 2.83 | 0.34 | + | IR,Mrk |
| 454 | ESO465-009 | 213515.1−305041 | 3195 | 59 | 5 | 12.32 | 1.66 | 0.41 | | |
| 455 | 2MASX... | 213554.0−030853 | 3073 | 5 | 4 | 11.16 | 2.08 | 0.45 | | IR |
| 456 | UGC11782 | 213809.7+085741 | 1359 | 5 | 8 | 11.71 | 2.37 | 1.05 | + | |
| 457 | 6dF... | 214226.9−061954 | 1448 | 74 | 9 | 13.36 | 1.96 | 0.64 | | |
| 458 | CGCG402-023 | 215102.8+061436 | 1698 | 28 | 9 | 12.77 | 2.60 | 0.16 | + | KIG |
| 459 | ESO532-002 | 215322.1−263441 | 1858 | 5 | 10 | 15.98 | 1.98 | 0.64 | + | |
| 460 | UGC11908 | 220635.2+160319 | 2039 | 5 | 9 | 13.84 | 1.89 | 0.58 | + | |
| 461 | UGC11921 | 220915.6+142136 | 1939 | 5 | 8 | 13.34 | 1.86 | 1.18 | + | KIG |
| 462 | ESO404-037 | 221035.2−365031 | 2610 | 5 | 8 | 14.19 | 1.88 | 1.34 | + | |
| 463 | APMBGC... | 221848.9−461303 | 1185 | 45 | 9 | 13.96 | 1.65 | | | |
| 464 | IC5201 | 222057.4−460209 | 893 | 5 | 6 | 10.16 | 1.65 | 1.97 | + | IR |
| 465 | ESO238-005 | 222230.1−482418 | 671 | 9 | 10 | 13.19 | 1.99 | 1.18 | + | |
| 466 | DUKST405-048 | 222310.0−333953 | 2512 | 74 | 9 | 13.94 | 1.65 | 0.3 : | + | |
| 467 | UGC12010 | 222318.7+053202 | 3037 | 5 | 5 | 11.37 | 2.56 | 0.70 | + | F |
| 468 | ESO289-026 | 222333.2−421628 | 2421 | 5 | 8 | 11.93 | 1.85 | 1.35 | | IR |
| 469 | ESO190-011 | 222703.3−521536 | 2788 | 9 | 8 | 12.26 | 1.89 | 0.58 | + | |
| 470 | ESO345-021 | 223237.8−380254 | 3193 | 49 | 4 | 12.20 | 2.15 | 0.5 : | | |
| 471 | ESO238-018 | 223543.8−505326 | 2848 | 75 | 5 | 12.39 | 1.89 | 0.70 | + | IR |
| 472 | DDO214 | 223635.0−025424 | 1887 | 5 | 8 | 10.56 | 1.93 | 1.08 | + | |
| 473 | NGC7328 | 223729.3+103154 | 3054 | 21 | 5 | 9.60 | 1.75 | 1.05 | + | IR,KIG |
| 474 | UGC12141 | 223814.8+804038 | 2353 | 5 | 7 | 11.80 | 1.62 | 0.92 | + | F,KIG |
| 475 | UGC12151 | 224133.9+002404 | 1966 | 5 | 8 | 11.27 | 1.79 | 1.20 | + | |
| 476 | SDSS... | 224359.9−100701 | 1115 | 6 | 10 | 15.03 | 2.78 | 0.3 : | + | |
| 477 | PGC133414 | 224431.4−281547 | 2604 | 21 | 9 | 13.68 | 1.98 | 0.5 : | | |
| 478 | UGC12178 | 224508.7+062551 | 2160 | 5 | 8 | 10.52 | 1.83 | 1.37 | + | IR,KIG |
| 479 | HIPASSJ2250+00 | 225024.2+005247 | 1902 | 9 | 10 | 14.43 | 1.79 | 0.5 : | + | |
| 480 | UGC12221 | 225025.6+825237 | 2314 | 11 | 7 | 11.07 | 1.62 | 1.14 | + | IR,KIG |

| (1) | (2) | (3) | (4) | (5) | (6) | (7) | (8) | (9) | (10) |
|---|---|---|---|---|---|---|---|---|---|
| 481 | UGC12213 | 225103.0+071745 | 3438 | 5 | 7 | 12.47 | 1.73 | 0.92 + | |
| 482 | NGC7406 | 225356.2−063445 | 1786 | 8 | 4 | 10.41 | 2.21 | 0.41 | |
| 483 | MCG-05-54-004 | 225445.2−265325 | 868 | 37 | 9 | 13.31 | 2.17 | 1.01 | IR |
| 484 | ESO469-006 | 225508.0−305520 | 3027 | 53 | −1 | 11.41 | 2.05 | 0.5 : | IR |
| 485 | LSBCF469-02 | 225721.5+275852 | 3233 | 6 | 8 | 15.67 | 2.93 | 0.42 + | |
| 486 | ESO603-031 | 225726.4−175308 | 2391 | 10 | 2 | 12.22 | 2.31 | 0.20 + | |
| 487 | PGC938611 | 225946.2−132322 | 1361 | 9 | 9 | 12.57 | 2.41 | 0.37 | |
| 488 | NGC7460 | 230142.9+021549 | 3401 | 18 | 5 | 9.95 | 1.77 | 0.07 + | IR |
| 489 | PGC132632 | 230313.0−352415 | 2157 | 43 | 6 | 13.94 | 1.63 | 0.48 + | |
| 490 | PGC141060 | 230341.3+234108 | 1432 | 6 | 10 | 14.13 | 2.67 | 0.50 | |
| 491 | UGC12340 | 230433.9+270921 | 1340 | 8 | 8 | 12.63 | 2.67 | 0.87 + | |
| 492 | ESO604-004 | 230628.9−174710 | 3120 | 5 | 8 | 13.72 | 2.56 | 0.94 + | |
| 493 | HIPASSJ2307-61 | 230720.8−614052 | 3108 | 9 | 10 | 15.06 | 1.74 | 0.41 + | |
| 494 | 2MASX... | 231122.0−231525 | 3490 | 74 | 3 | 13.00 | 2.77 | 0.7 : | |
| 495 | MCG-03-59-001 | 231136.7−152835 | 2201 | 20 | 7 | 12.12 | 1.72 | 0.93 + | F |
| 496 | ESO407-014 | 231739.5−344727 | 2782 | 7 | 5 | 10.80 | 2.15 | 1.03 + | IR |
| 497 | 6dF... | 231803.9−485936 | 2275 | 74 | 9 | 14.30 | 1.67 | 0.3 : | |
| 498 | UGC12495 | 231848.5+163738 | 2869 | 21 | 6 | 13.71 | 1.87 | 0.66 + | |
| 499 | KUG2316+162 | 231921.9+163401 | 3187 | 18 | 5 | 13.34 | 1.86 | −0.07 + | |
| 500 | KKR75 | 232011.2+103723 | 1703 | 5 | 10 | 15.44 | 1.62 | 0.54 + | |
| 501 | FGC2498 | 232442.0−024730 | 2726 | 13 | 7 | 13.76 | 1.96 | 0.3 : | F |
| 502 | IC5321 | 232620.0−175723 | 2984 | 11 | 5 | 12.29 | 2.29 | 1.01 + | IR |
| 503 | UGC12070 | 233157.3+780903 | 1724 | 5 | 10 | 13.24 | 2.10 | 0.96 + | |
| 504 | ESO347-029 | 233627.2−384657 | 1553 | 8 | 8 | 11.28 | 1.74 | 1.46 + | |
| 505 | UGCA441 | 233739.6+300746 | 1660 | 7 | 9 | 12.06 | 2.89 | −0.14 + | IR,Mrk,KIG |
| 506 | UGC12713 | 233814.7+304233 | 568 | 8 | 9 | 13.60 | 2.09 | 0.88 + | KIG |
| 507 | UGC12729 | 234020.8+011445 | 2074 | 55 | 4 | 12.24 | 2.21 | 0.98 + | KIG |
| 508 | 2MASX... | 234056.8−190507 | 1637 | 30 | 9 | 12.35 | 2.12 | 0.34 | IR |
| 509 | ESO292-014 | 234235.2−445417 | 1500 | 55 | 6 | 10.75 | 1.68 | 1.06 + | IR,F |
| 510 | PGC072271 | 234425.9−164849 | 1680 | 9 | 10 | 13.36 | 2.12 | 0.73 + | |
| 511 | HIPASSJ2344-07 | 234429.5−073545 | 1983 | 9 | 8 | 13.25 | 2.52 | 0.76 | |
| 512 | UGC12771 | 234532.7+171512 | 1535 | 5 | 10 | 13.90 | 3.25 | 0.46 + | |
| 513 | ESO149-001 | 234749.9−570415 | 1800 | 11 | 8 | 9.49 | 1.87 | 1.82 + | IR |
| 514 | PGC812517 | 234951.8−223256 | 1098 | 9 | 9 | 13.69 | 2.88 | 0.46 + | |
| 515 | NGC7764 | 235054.0−404342 | 1691 | 69 | 8 | 10.30 | 1.74 | 1.05 + | |
| 516 | ESO149-003 | 235202.8−523440 | 501 | 9 | 12 | 12.41 | 2.48 | 1.15 + | |
| 517 | UGC12856 | 235645.3+164850 | 2017 | 8 | 8 | 13.21 | 1.67 | 1.39 + | pec |
| 518 | UGC12857 | 235647.6+012118 | 2642 | 7 | 5 | 11.26 | 1.75 | 1.36 + | IR,F,KIG |
| 519 | ESO028-007 | 235801.0−732711 | 2374 | 9 | 7 | 13.00 | 3.46 | 0.86 | |
| 520 | NGC7800 | 235936.7+144826 | 1976 | 6 | 10 | 11.26 | 1.61 | 1.71 + | IR,pec |

The least frequent among the isolated galaxies are the lenticular and Sa–Sb galaxies ($T = 0 − 3$). In the KIG and 2MIG catalogs of isolated galaxies the median of the distribution by type lies between $T = 3$ (Sb) and $T = 4$ (Sbc), i.e. a half of the population of these samples has well-defined bulges. The relative number of irregular galaxies in the KIG and 2MIG catalogs is less than 4%.

However, a considerable part of differences in the morphological composition of the LOG, KIG and 2MIG catalogs is due to the selection effects. As a distance-limited sample, the LOG catalog contains much more galaxies of low luminosity ($T = 10, 9$) than the KIG and 2MIG catalogs, where the selection of galaxies was based on their apparent magnitude, i.e. luminosity. We observe a similar effect on the example of galaxies in the Local Volume with $D < 10$ Mpc, where about 70% of the total sample belong to the types 9 and 10, i.e. to irregular galaxies. We also note that among

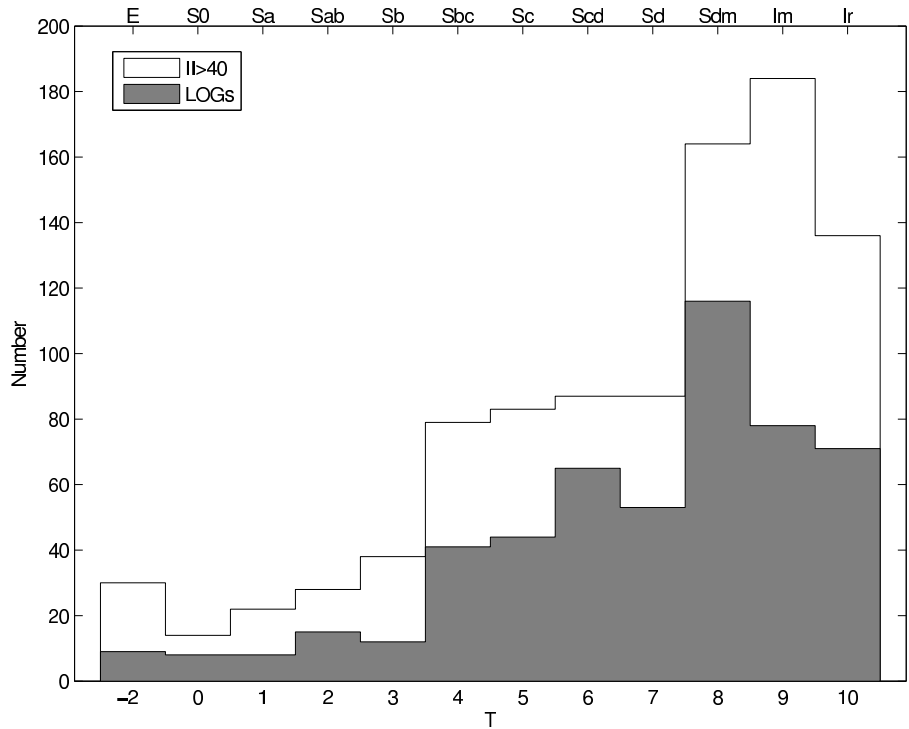

**Figure 4.** The distribution of isolated galaxies by morphological type. The upper histogram shows all the 990 galaxies with $II > 40$, the gray histogram—520 galaxies from the LOG catalog.

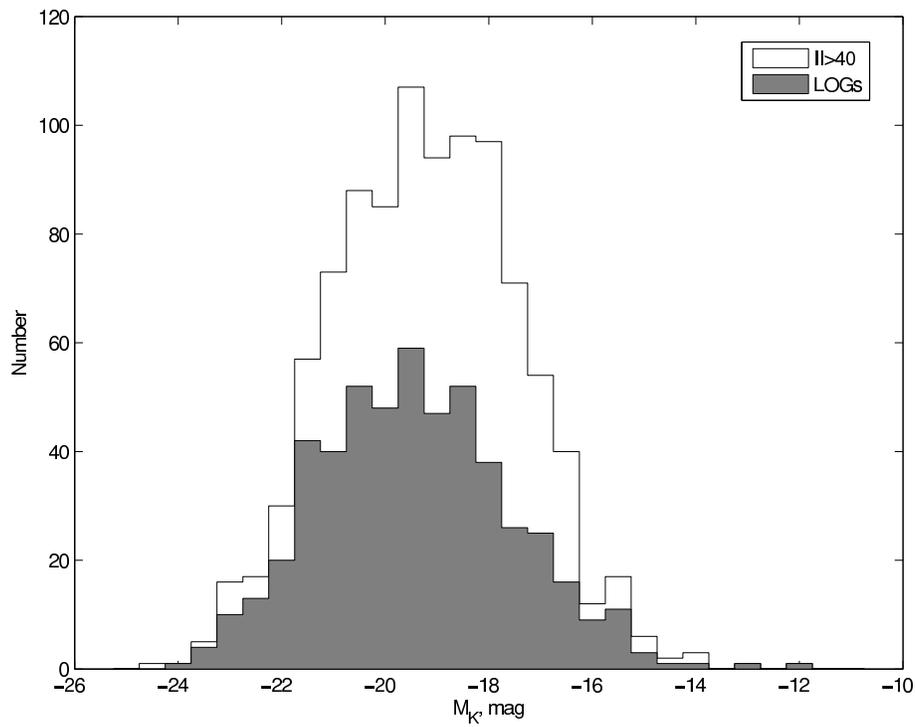

**Figure 5.** Luminosity function of isolated galaxies. Gray marks the catalog distribution of the LOG galaxies.

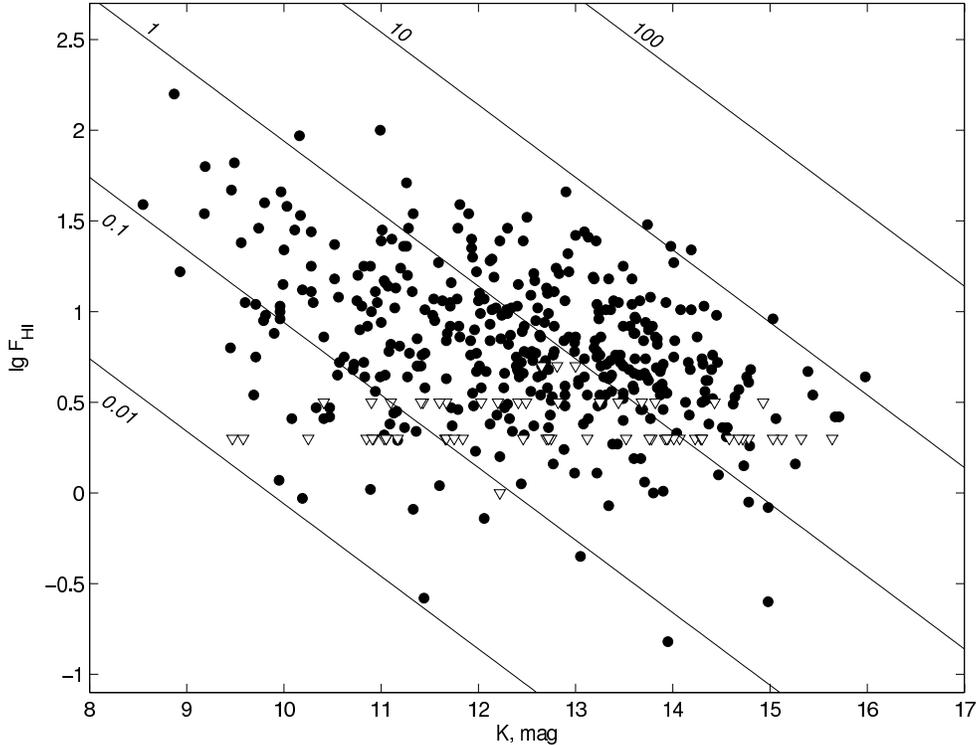

**Figure 6.** The apparent $K$ magnitudes and the HI flux logarithm for the LOG galaxies. The triangles show the upper values of the HI flux. The sloping lines correspond to the constant ratio $M_{HI}/L_K$, equal to 0.01, 0.1, 1, 10 and 100 in solar units.

33 isolated galaxies, common to the KIG and LOG catalogs, the peak of the distributions by the morphological type falls on $T = 6$ (Scd), i.e. on the spirals without apparent bulges.

It must be emphasized that it is usually difficult to find the evolutionary reasons for the differences of two samples by morphological composition, since these differences are masked by the selection effects of the sample itself. For example, two catalog samples of isolated galaxies, KIG and 2MIG, show an apparent difference in the relative number of galaxies with bulges. However, this is an expected difference, and it is due to the fact that the galaxies in the 2MIG catalog were selected by luminosity in the near infrared, where the galaxy bulges are more prominent than in the optical $B$-band.

As follows from the last column of Table 1, about 30% of the LOG galaxies are the infrared IRAS-sources. Their distribution by morphological type varies considerably from the total distribution (see Fig. 4 in [14]). The maximum of the distribution of isolated IRAS galaxies falls on the $T = 4-6$ types. Such a difference with the total sample has a highly selective character, since the IRAS fluxes are closely correlated with the mass of the dust component of galaxies, which is most pronounced

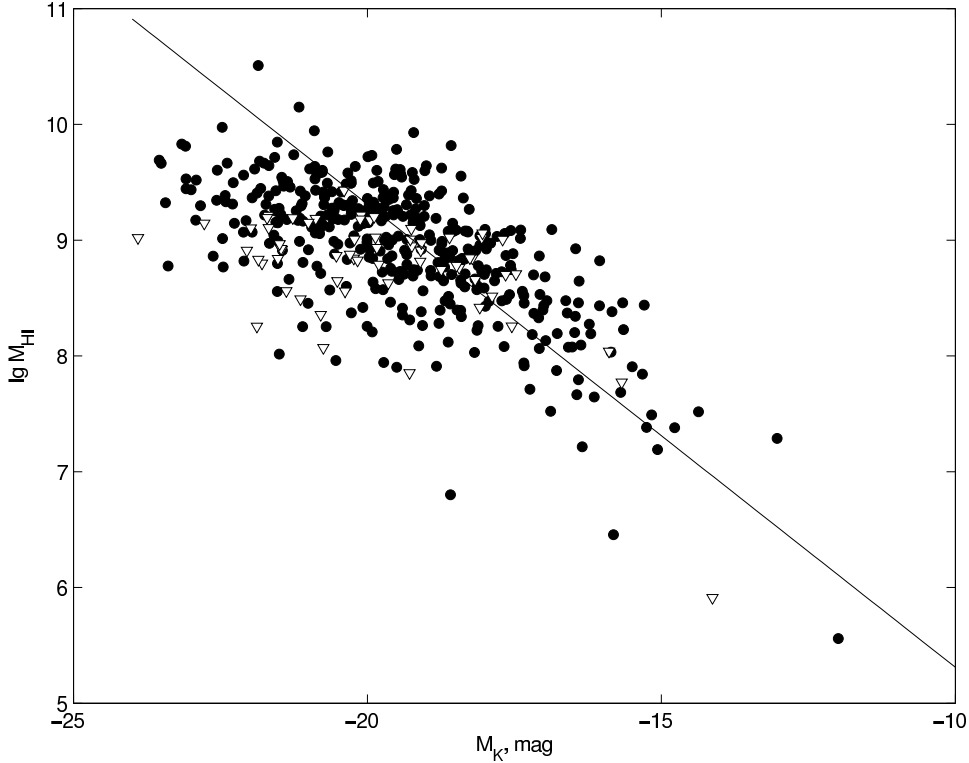

**Figure 7.** The logarithm of the hydrogen mass vs. the absolute $K$ magnitude in LOG galaxies. The solid line corresponds to the case $M_{gas} = M_{stars}$ in the galaxy.

namely in the Sc galaxies.

Current models of galaxy formation imply that the E and S0 galaxies are formed by multiple mergers of disk-shaped and irregular galaxies. Clusters usually involve a permanent mechanism of transformation of spiral galaxies into lenticular by sweeping out the gas, hence preventing further star formation. Thus, it is expected that the E and S0 galaxies should occur mainly in the regions of high matter density. Indeed, the KIG and 2MIG catalogs contain only 16% and 18% of the E + S0 galaxies, respectively, which is lower than their average cosmic abundance (about 24%).

There are only 17 galaxies in the LOG catalog that we classified as E and S0 (T<1). A list of these galaxies, making up less than 4% of our sample is presented in Table 2. Hereinafter the distances to the galaxies were determined from their radial velocity $V_{LG}$ at the Hubble constant $H_0 = 73$ km/s Mpc$^{-1}$.

It is noteworthy that the isolated E and S0 galaxies in the LOG differ significantly from the other E and S0 galaxies by a number of features. The isolated early-type galaxies have a rather low luminosity, their median absolute magnitude is $M_K = -21.7$ or $M_B = -17.7$. More than a half of E and S0 galaxies in the LOG are the IRAS sources, which indicates the presence of a considerable amount of dust. Some of them (NGC 404, UGC 1198, UGC 5467) reveal the HI line fluxes, corresponding to the masses of neutral hydrogen of the order of $10^8 - 10^9 M_\odot$. We can assume that during their evolution the isolated E and S0 galaxies undergo a significant impact (inflow) of the intergalactic medium, where about 80% of the universe's baryons are concentrated [15].

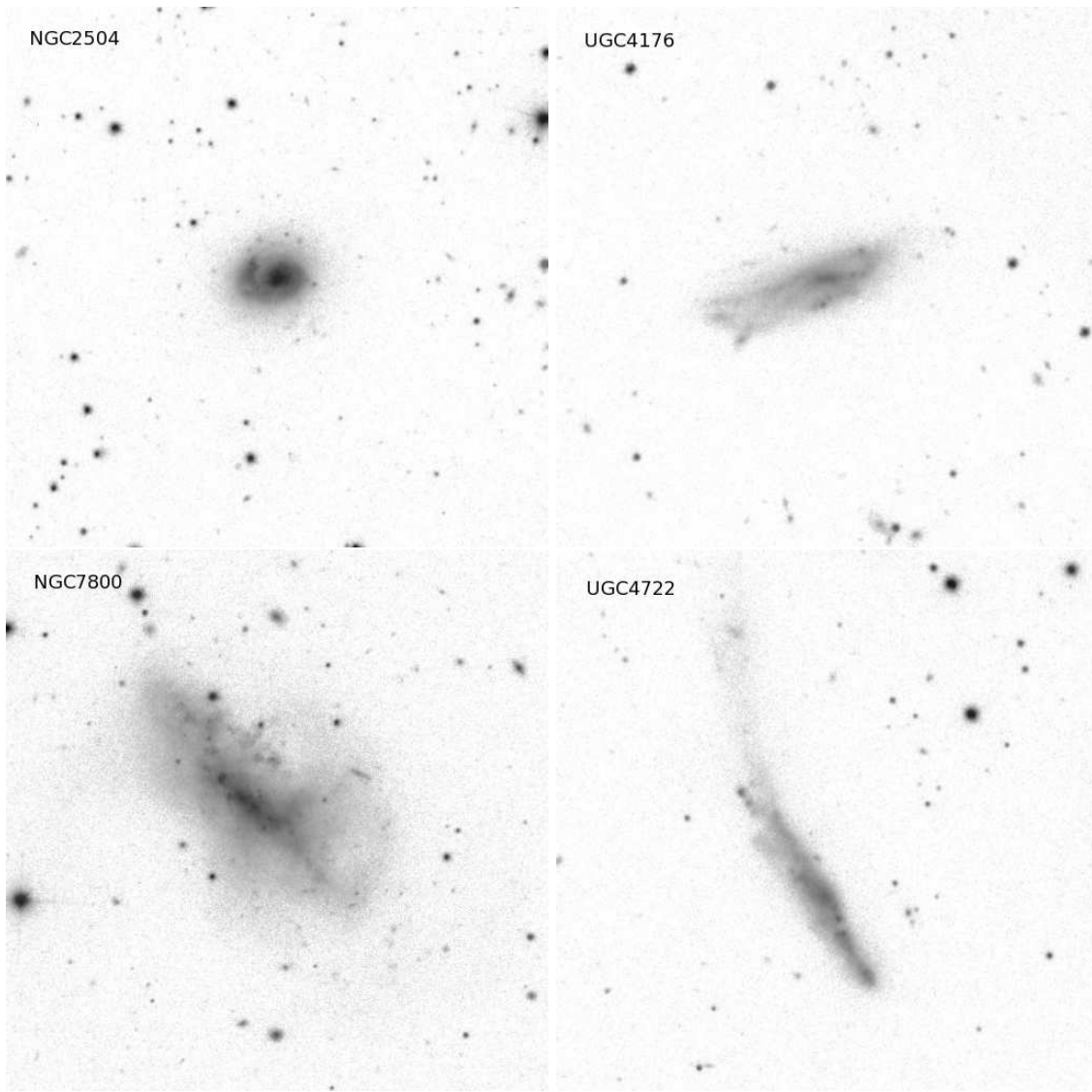

**Figure 8.** Examples of isolated galaxies with peculiar structures: NGC 2504, UGC 4176, NGC 7800, and UGC 4722.

Figure 5 demonstrates the distribution of all types of isolated galaxies by absolute $K$ magnitudes. The galaxies that are included in the LOG catalog (in gray) have approximately the same luminosity function as all the 990 galaxies satisfying the condition $II > 40$. In other words, an additional rejection of isolated galaxies using Karachentseva's criterion does not introduce any significant selection by luminosity.

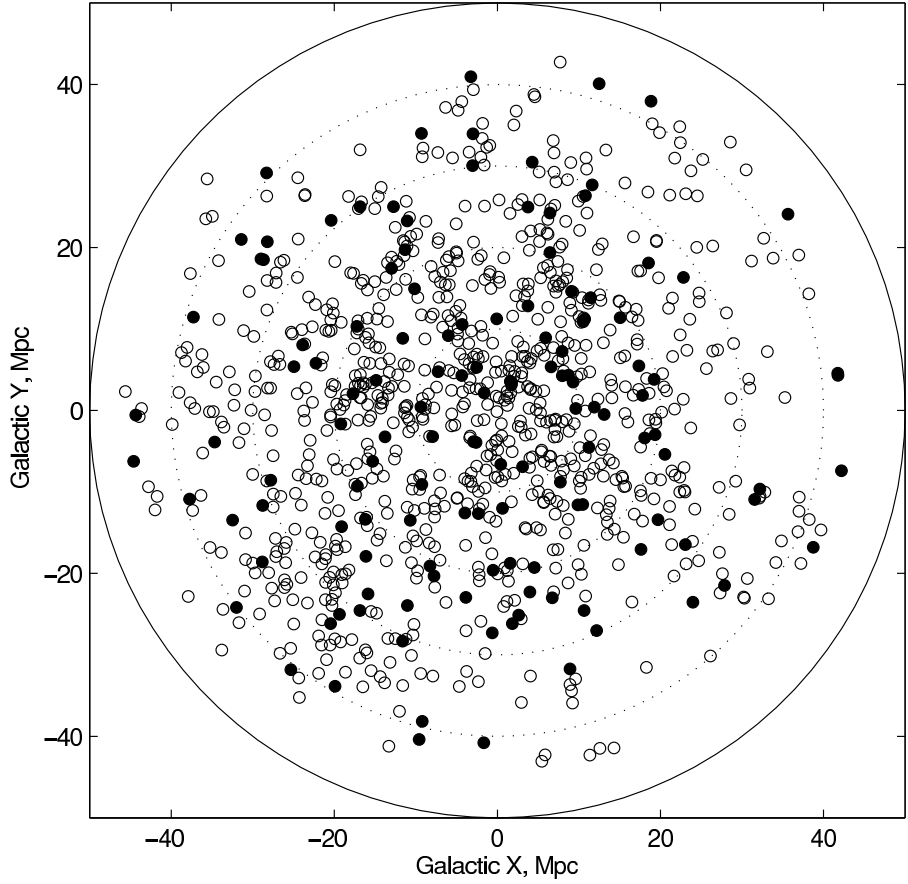

**Figure 9.** The distribution of 520 LOG galaxies in Cartesian Galactic coordinates. Galaxies with developed bulges (T< 4) are shown by solid circles.

**Table 2.** Galaxies of early morphological types (T < 1) in the LOG catalog

| LOG | $V_{LG}$ | $T$ | $M_K$ | lg $M_{HI}$ | Note |
|---:|---:|---:|---:|---:|:---|
| 31 | 3066 | 0 | −21.5 | <8.9 | IRAS |
| 50 | 218 | −3 | −18.9 | 7.9 | IRAS |
| 62 | 1408 | −2 | −20.6 | 8.0 | IRAS |
| 104 | 3449 | −1 | −23.9 | <9.0 | IRAS |
| 169 | 2796 | 0 | −22.1 | − | |
| 220 | 3043 | 0 | −22.1 | <8.9 | |
| 231 | 2768 | −2 | −22.1 | 9.2 | IRAS |
| 256 | 1876 | 0 | −19.0 | − | |
| 264 | 2226 | −3 | −21.6 | <8.8 | |
| 275 | 3249 | −2 | −21.5 | <9.0 | |
| 332 | 1782 | 0 | −21.7 | − | IRAS |
| 407 | 2120 | −1 | −20.3 | − | |
| 411 | 3431 | 0 | −21.4 | − | IRAS |
| 420 | 3205 | 0 | −23.1 | − | |
| 435 | 2692 | −1 | −22.2 | − | IRAS |
| 450 | 3164 | 0 | −22.8 | <9.1 | IRAS |
| 484 | 3027 | −1 | −21.7 | <9.1 | IRAS |
| Median | | | −21.7 | <8.9 | |

**Table 3.** Nearby isolated galaxies with peculiar structures

| LOG | $V_{LG}$ | $T$ | $M_K$ | Structural features |
|---:|---:|---:|---:|---|
| 12 | 1636 | 2 | –19.6 | a weak external ring |
| 58 | 677 | 10 | –16.2 | dIr + cirrus? |
| 100 | 2777 | 8 | –19.2 | a wide curved tail |
| 114 | 2327 | 8 | –19.8 | flocky |
| 163 | 964 | 10 | –16.9 | flocky |
| 179 | 2419 | 4 | –21.2 | tail or arm from a compact body |
| 180 | 3092 | 7 | –20.5 | asymmetrical |
| 197 | 1705 | 7 | –19.2 | a wide tail |
| 219 | 1811 | 9 | –18.7 | knotted with a tail |
| 243 | 2377 | 8 | –19.5 | knotted |
| 260 | 2635 | 10 | –20.3 | knotted |
| 273 | 2141 | 10 | –18.0 | knotted |
| 302 | 1246 | 8 | –18.2 | hammer-like |
| 304 | 939 | 8 | –19.4 | faint blue knots in the outskirts |
| 310 | 2331 | 2 | –19.0 | a weak extended periphery as in Malin-1 |
| 338 | 1477 | 9 | –17.9 | comet-like |
| 340 | 3350 | 7 | –19.5 | asymmetrical |
| 347 | 3469 | 4 | –21.7 | a distorted tail |
| 447 | 1313 | 8 | –18.7 | asymmetrical, flocky |
| 517 | 2017 | 8 | –19.0 | distorted, flocky |
| 520 | 1976 | 10 | –20.9 | disturbed, with loops |
| Median | | 8 | –19.2 | |

Figure 6 presents the distribution of LOG galaxies by apparent $K$ magnitudes and by the logarithms of the integrated HI line flux. Each galaxy with measured HI flux is marked by a circle, and the galaxies with an estimate of the upper limit of HI flux are marked by triangles. Apart from them, about 12% of galaxies in this catalog have not yet been observed in the HI line. The diagonal lines in the figure correspond to the values of the hydrogen mass-to-stellar mass ratio, equal to 0.01, 0.1, 1, 10 and 100. Here the mass ratio was defined as

$$\log(M_{HI}/M_*) = \log F_{HI} + 0.4 m_K - 5.94$$

at the absolute $K$-luminosity of the Sun $M_K(\odot) = 3.28^m$. These data show that in almost all the galaxies the $M_{HI}/M_*$ ratio lies within the range from 0.01 to 10 with a median value of 0.7. Taking into account the correction for helium and molecular hydrogen (the factor of 1.85 according to [15]), the characteristic ratio of the mass of gas and stars in the considered galaxies is equal to 1.3. Hence, a typical isolated LOG galaxy has transformed into stars less than a half of its gas reserves. And in some isolated galaxies (UGC 3672) more than 90% of baryons still reside in the form of the gas component.

It is well known that the mass ratio of gas to stars shows a tendency of increase from normal to dwarf galaxies. This could mean that the phase of active star formation in dwarf galaxies comes later than normal, or that the star formation in dwarf systems advances in a more slow rate. Figure 7 reproduces the relationship between the logarithm of hydrogen mass and the absolute $K$ magnitude of isolated galaxies. The objects with an estimate of the upper limit of HI flux are marked by triangles. The straight line in the figure fixes the value of $M_{HI} = M_*$. The galaxies with higher luminosity ($M_K < -19.5^m$) show a significant hydrogen mass deficiency with respect to this line.

## 5. PECULIAR ISOLATED GALAXIES

The standard cosmological $\Lambda$CDM model predicts the existence of a large number of massive invisible bodies (dark sub-haloes), in which the star formation process has not yet been triggered [16]. It is assumed that such dark clumps with masses on the order of $(10^6 - 10^9)M_\odot$ can ten times more numerous than normal galaxies. The search for such objects in the HIPASS and ALFALFA surveys via the HI line emission were so far unsuccessful [17, 18].

The presence of a numerous population of dark massive objects among the galaxies should give a large number of tight interactions "galaxy–dark body." The expected result of their close approach could be a distortion of the structure of a normal (bright) galaxy, or a "visualization" of a dark component via an inflow of matter from the ordinary galaxy. It is obvious that the cases of "unmotivated" interaction with an invisible object are to be searched just among the isolated galaxies. Karachentsev et al. [19, 20] have found 8 examples of isolated galaxies from the KIG sample with pronounced structural distortions. Peculiar shapes that some of them have may be due to an asymmetric starburst or a recent merger of two galaxies. However, two galaxies: UGC 4722 and ESO 545–05 proved to match quite adequately the assumption about their interaction with a massive invisible object. The interaction effects are most pronounced in the cases where the objects of approximately the same mass approach each other. Since it is expected that the number of dark bodies increases with the decreasing mass, the search for traces of interaction with such bodies looks more promising among the most low-mass isolated galaxies.

A typical galaxy in the LOG catalog has on the average an order of magnitude lower luminosity (mass) than a typical representative of the KIG catalog. The last column of Table 1 lists 21 galaxies with a peculiar morphology. All of them are enumerated in Table 3. The reproductions of four interacting isolated galaxies from the SDSS and DSS surveys are demonstrated in Fig. 8. The existence of a very asymmetric shape (NGC 2504), a distorted spiral structure (NGC 4176), a perturbed disk with loops (NGC 7800), a broad tidal tail (UGC 4722) is rather difficult to explain without invoking the idea of interaction with an invisible massive body. However, it should be emphasized that the reported cases of peculiar structures make up only 4% of the total number of galaxies in the LOG catalog. This fact can become a strong restriction on the size of the population of dark objects with masses of about $10^8 - 10^9 M_\odot$.

## 6. DISCUSSION AND CONCLUSIONS

The above data show that the list of most isolated galaxies is dominated by the objects of late morphological types and low luminosity, most of which possess significant reserves of gas for the further star formation. These properties seem to be quite anticipated in the framework of the standard cosmological model. As Makarov and Karachentsev have noted [21], about 46% of galaxies are located outside the virial regions of groups and clusters, however only 18% of the total stellar mass (i.e. the luminosity of galaxies in the $K$-band) is located outside the groups and clusters.

All the 22 LOG catalog galaxies, located in the Local Volume of the 10 Mpc radius have a negative value of the tidal index (3). One of the closest and most isolated LOG galaxies in the Local Volume is a well-known irregular galaxy KK 246, located on the outskirts of the Local cosmic Void. The median value of the tidal index for these 22 galaxies, $TI = -1.7$, shows that the typical local density around the LOG galaxies is approximately 50 times lower than the average density in the Local Volume. This fact emphasizes the physical basis for the isolation criteria used in our catalog.

The number of galaxies in the LOG, common with the KIG catalog amounts to mere 33. A small fraction of the overlap between the LOG and KIG catalogs is due primarily to the difference in spatial volumes of these samples, which overlap by about $1/8$. A detailed comparison of galaxy properties in both catalogs deserves special consideration.

What stands out is a relatively high number of "flat" spiral galaxies, seen edge-on ($N_F = 71$ in the last column of Table 1). Taken a random orientation of axes in thin spiral galaxies, their expected number in the LOG catalog should be about 18–32. The observed excess of the $N_F$ number indicates that the thin disks with an axis ratio $a/b > 7$ can persist mainly in the sites of low matter density, where the tidal perturbations of neighbors do not cause the "warming-up" and thickening of the disc along the axis of rotation.

Quite a large number of galaxies in the LOG are the IRAS sources ($N_{IRAS} = 142$). Their distribution by morphological types is characterized by a broad maximum at the Sbc–Scd types ($T = 4 - 6$), what indicates the presence of a developed dust component in the disks of isolated galaxies. The total number of "active" galaxies (Mrk) in the LOG catalog is relatively small (18), most of them being intergalactic HII regions [22], rather than objects with active (Seyfert) cores.

Figure 9 demonstrates the distribution of 520 LOG galaxies in Cartesian Galactic coordinates in the projection on the plane of the Milky Way. The choice of such a projection reduces the effect of absorption of light in our Galaxy. In the first approximation, the distribution of isolated galaxies looks quite homogeneous, with some decrease in the number density of galaxies from the center to the periphery of the volume due to the loss of a part of dwarf galaxies at large distances from the observer. The galaxies with bulges ($T < 4$), marked by filled circles, do not reveal any increased crowding in groups compared with the later types (empty circles). The fact that the isolated galaxies avoid the volumes occupied by nearby clusters and groups, as well as nearby cosmic voids (the Local Void in the Aquila-Hercules, in the Eridanus and in the Leo) has some influence on their spatial distribution [23].

At the same time, one can recognize a slightly pronounced mutual association of the LOG galaxies with radial velocity differences of less than 50 km/s on a scale of around 0.5 Mpc. Examples of such associations are LOG 254+255, LOG 360+362, LOG 3+20+30. The presence of such non-virialized structures may indicate the existence in the regions of low density of some low-contrast filamentary structures (similar to the nearby chain in Sculptor [24]) or an association of dwarf galaxies, noted recently in [25] and [3]. The features of galaxy distribution in the regions of extremely low matter density contain important information on the formation and evolution of the large-scale structure of the universe. However, this sector of observational cosmology is still almost unexplored.

## ACKNOWLEDGMENTS

This study was made owing to the support of the following grants: the grants of the Russian Foundation for Basic Research (RFBR) nos. 09–02–90414-UKR-f-a, 08–02–00627-a, and the grant of the Ministry of Education and Science of the Ukraine no. F28.2/059. The paper made use of the HyperLEDA (`http://leda.univ-lyon1.fr`) and NED (`http://nedwww.ipac.caltech.edu`) data- bases.

---

1. R.B. Tully, Astrophys. J. **321**, 280 (1987)

2. A.C. Crook, J.P. Huchra, N. Martimbeau, et al., Astrophys. J. **655**, 790 (2007).

3. D.I. Makarov and I.D. Karachentsev, Monthly Notices Roy. Astronom. Soc. (2011) (accepted).

4. V.E. Karachentseva, Soobscheniya SAO **8**, 3 (1973).

5. V.E. Karachentseva, S.N. Mitronova, O.V. Melnyk, and I.D. Karachentsev, Astrophysical Bulletin **65**, 1 (2010).

6. T.H. Jarrett, T. Chester, R. Cutri, et al., Astronom. J. **119**, 2498 (2000).

7. I.D. Karachentsev, V.E. Karachentseva, W.K. Huchtmeier, and D.I. Makarov, Astronom. J. **127**, 2031 (2004).

8. G. Paturel, C. Petit, Ph. Prugniel, et al. Astronom. and Astrophys. **412**, 45 (2003).

9. T. Jarrett, R. Chester, R. Cutri, et al, Astronom. J. **125**, 525 (2003).

10. I.D. Karachentsev and A.M. Kut'kin, Astron. Lett. **31**, 299 (2005).

11. O.V. Melnyk, V.E. Karachentseva, I.D. Karachentsev, et al., Astrofizica **52**, 184 (2009).

12. G. de Vaucouleurs, A. de Vaucouleurs, H.G. Corwin, et al., Third Reference Catalogue of Bright Galaxies, series I-III, Springer-Verlag Berlin Heidelberg New York, (1991).

13. S.S. Alam, D.L. Tucker, B.C. Lee, and J.A. Smith, Astronom. J. **129**, 2062 (2005).

14. I.D. Karachentsev, D.I. Makarov, V.E. Karachentseva, and O.V. Melnyk, ASP Conference ser. **421**, 69 (2010).

15. M. Fukugita and P.J.E. Peebles,Astrophys. J. **616**, 643 (2004).

16. A.A. Klypin, A.V. Kravtsov, and O. Valenzuela, Astrophys. J. **522**, 82 (1999).

17. J.I. Davies, Proceedings IAU Symp. №244, 7 (2008)

18. M.P. Haynes, Proceedings IAU Symp. №244, 83 (2008).

19. I.D. Karachentsev, A.E. Dolphin, R.B. Tully, et al., Astronom. J. **131**, 1361 (2006).

20. I.D. Karachentsev, V.E. Karachentseva, W.K. Huchtmeier, et al., Proceedings IAU Symp. №244, 235 (2008).

21. D.I. Makarov and I.D. Karachentsev, Astrophysical Bulletin **64**, 24 (2009).

22. L. Searle and W.L.W. Sargent, Astrophys. J. **173**, 25 (1972)

23. A.V. Tikhonov and I.D. Karachentsev, Astrophys. J. **653**, 969 (2006).

24. I.D. Karachentsev, E.K. Grebel, M.E. Sharina, et al., Astronom. and Astrophys. **404**, 93 (2003).

25. R.B. Tully, L. Rizzi, A.E. Dolphin, et al., Astronom. J. **132**, 729 (2006).